\documentstyle[aps,multicol,psfig]{revtex}
\begin{document}
\draft

\newcommand{\bd}[1]{ \mbox{\boldmath $#1$}  }

\title{Electron scattering on two--neutron halo nuclei: The case of $^6$He}
\author{E. Garrido \and E. Moya de Guerra}
\address{Instituto de Estructura de la Materia, CSIC, Serrano 123, E-28006
Madrid, Spain}
\date{\today}

\maketitle

\begin{abstract}

The formalism to describe electron scattering reactions on two--neutron
halo nuclei is developed. The halo nucleus is described as a three--body
system (core+n+n), and the wave function is obtained by solving the Faddeev
equations in coordinate space. We discuss elastic and quasielastic scattering 
using the impulse approximation to
describe the reaction mechanism. We apply the method to investigate the
case of electron scattering on $^6$He. Spectral functions, response functions, 
and differential cross sections are calculated for both neutron knockout 
and $\alpha$ knockout by the electron.


\end{abstract}

\begin{multicols}{2}

\section{Introduction}

In the middle of the 80's it was experimentally found that some light 
nuclei close to the neutron drip line have an spatial extension much
larger than expected according to its mass number 
\cite{tan85,tan85b,kob88,tan88}. This is the case for instance of $^{11}$Li, 
that has a root mean square radius similar to a nucleus with a mass number 
three times larger. 
Very soon it was suggested that these nuclei could be understood as a
core surrounded by one or more  neutrons, which extend to several times the
nuclear radius \cite{han88}. This picture was supported by subsequent
studies that proved the validity of the few--body models as an appropriate
method to describe the most general properties of halo nuclei 
\cite{tos90,joh90,zhu91,zhu91b,rii92}.

The term halo nuclei has then been coined to describe very  weakly bound 
and spatially extended systems, 
where several (usually one or two)  neutrons have a high probability of 
being at distances larger than the typical nuclear radius. 
For a general overview of their basic properties see for instance
\cite{han95}.
Among halo nuclei a lot of attention has been paid to two--neutron halo 
systems from which $^6$He ($\alpha+n+n$) and $^{11}$Li ($^9$Li $+n+n$) 
are their maximum exponents 
\cite{zhu93}.  They are samples of the so called Borromean nuclei, that 
have the property of being bound while  all the two--body subsystems 
made by two of the three constituents are unbound.

The ability to produce secondary beams of halo nuclei opened the possibility
of investigating their structure by 
the measurement of the momentum distributions of the fragments
coming out after a collision with stable nuclei
\cite{ann90,orr92,orr95,zin95,nil95,hum95,zin97}.  The simplest approach
to the understanding of Borromean nuclei fragmentation reactions was
made by means of the sudden approximation \cite{zhu94,kor94,zhu95},
that assumes that  one of the three particles in the projectile is 
instantaneously removed by the target, while the other two particles 
remain undisturbed. Clearly this can only be justified for reaction times
much shorter than the characteristic time for the motion of the 
three particles in the system. Since the system is weakly bound this
requirement is well fulfilled for a high energy beam. This model was 
proved to be
valid to describe the core momentum distributions \cite{zhu95},
but it failed in the attempt of reproducing experimental momentum
distributions for neutrons. Several authors suggested that the final
interaction between the two non--disturbed particles played an essential
role, especially when low--lying resonances are present
\cite{zin95,kor94,bar93}. Incorporation into the model of the final 
state interaction \cite{gar96,gar97} is in fact necessary in order to
obtain a good agreement between theory and experiment also for 
neutron momentum distributions. Indeed further refinements
\cite{gar97b,gar98} in the description of the reaction
process have been proved to be crucial for a better interpretation
of the experimental data.

Although a lot of information has been extracted from the 
halo nucleus--nucleus fragmentation reactions, it may pay to look
for cleaner ways of investigating the structure
of halo nuclei. As in any nucleus--nucleus collision 
the effects of the strong interaction involved in the reaction 
mechanism are mixed with those determining the nuclear structure.  
It is very well known that the way to avoid this problem is to 
substitute the hadron probe by electrons. In fact electron scattering
is usually considered as the most powerful tool to investigate 
nuclear structure \cite{wal95,kel96}. The electromagnetic coupling constant 
is small 
enough such that the Born approximation can be used, the electron--nucleus
interaction is well described through QED, and it is possible to 
vary independently the momentum transfer and the energy transfer
to the nucleus. The problem when dealing with halo nuclei is that
they are far from the stability region, and the nuclear target can
not be at rest as in conventional nuclear structure studies with electron
beams. However it is in principle possible to perform
electron scattering experiments by means of the collision between a
secondary beam of unstable nuclei and an electron beam. This kind
of experiments are part of the MUSES (MultiUSse Experimental Storage 
rings) project in RIKEN, and they are projected for the beginning of
the coming century \cite{ohk97}. 

In parallel with the experimental projects it is then clear that
theoretical studies of the process should be developed.
From the experimental point of view
the simplest reaction would be the elastic scattering. This would
permit to investigate the charge density of the halo nucleus.
This is specially interesting when the proton dripline is approached
because it directly gives the extension of the possible proton
halo ($^8$B is a good candidate for this). For neutron halos 
the charge radius is expected to be the one of the core. Elastic 
electron scattering on $^{11}$Li or $^6$He will be important to
confirm that this picture of neutron halo nuclei is correct. 
As one slightly increases the energy transfer to the nucleus the halo will
break up, and coincidence $(e,e^\prime x)$ measurements can be made,
with $x$ either a halo neutron or the stable core.

The main goal of this paper is to contribute to the  first
steps in the investigation of electron scattering reactions on two--neutron 
halo nuclei. To illustrate the model we will consider the case of
electron--$^6$He scattering. Although estimates of the elastic 
scattering reaction will be given, most of the work will be devoted 
to $(e,e^\prime n)$ and $(e, e^\prime \alpha)$ reactions at 
quasielastic kinematics.

The paper is organized as follows: A brief description of kinematics
and of elastic scattering reactions is given in sections II and III,
respectively. The differential cross section for exclusive 
$(e,e^\prime x)$ processes is given in section IV. Sections V
and VI are devoted to the theoretical description of the different
ingredients involved in the electron--$^6$He scattering process
and to the presentation of the results, respectively. 
Concluding remarks are given in section VII.  

\end{multicols}
\begin{figure}[ht]
\centerline{\psfig{figure=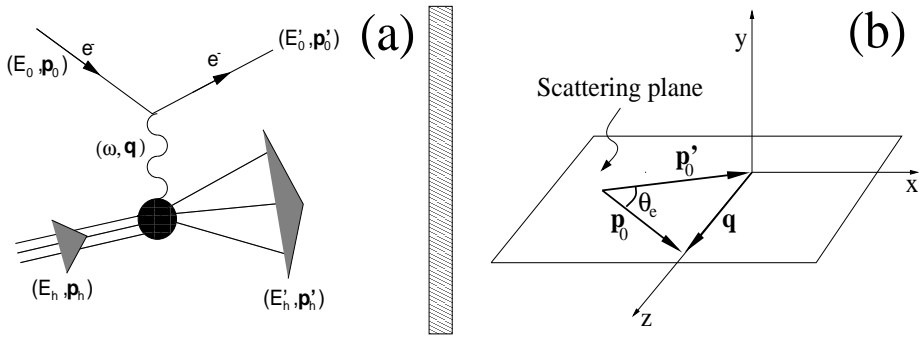,width=8.5cm,%
bbllx=6.5cm,bblly=6.5cm,bburx=15.2cm,bbury=21.2cm,angle=270}}
\vspace{0.2cm}
\caption[]{
(a) Scheme of a general electron--nucleus reaction in the
Born approximation. (b) Axes system chosen to describe the process.
}
\end{figure}
\begin{multicols}{2}

\section{Kinematics and general considerations}

In fig.1a we show the scheme of a general electron scattering reaction.
An electron with energy and momentum $(E_0,\bd{p}_0)$ hits a target
with energy and momentum $(E_h,\bd{p}_h)$. The energy and momentum transfer
to the nucleus is denoted by $(\omega,\bd{q})$, and the energy and momentum
of the scattered electron and the final nuclear system are given by
$(E^\prime_0,\bd{p}^\prime_0)$ and $(E^\prime_h,\bd{p}^\prime_h)$, respectively.

In fig.1b we show our axis system. The $z$--axis is chosen along the momentum
transfer $\bd{q}$. The $y$--axis is defined along 
$\bd{p}_0 \times \bd{p}^\prime_0$,
and it is therefore perpendicular to the scattering plane defined by
$\bd{p}_0$ and  $\bd{p}^\prime_0$. The $x$--axis is defined by 
$\hat{\bd{y}} \times \hat{\bd{z}}$, where $\hat{\bd{z}}$ and $\hat{\bd{y}}$
are unit vectors along the $z$--axis and $y$--axis, respectively.
The angle between $\bd{p}_0$ and $\bd{p}^\prime_0$ is the scattering
angle $\theta_e$.

Energy and momentum conservation in the process determine that
\begin{eqnarray}
E_0 + E_h & = & E^\prime_0 + E^\prime_h \nonumber \\
\bd{p}_0 + \bd{p}_h & = & \bd{p}^\prime_0 + \bd{p}^\prime_h
\label{eq1}
\end{eqnarray}
and the energy and momentum transfer are related to the electron and nuclear
energy and momentum by
\begin{eqnarray}
\omega & = & E_0 - E^\prime_0  =  E^\prime_h - E_h \nonumber \\
\bd{q} & = & \bd{p}_0 - \bd{p}^\prime_0 = \bd{p}^\prime_h - \bd{p}_h
\label{eq2}
\end{eqnarray}
 
Working in the frame of the nucleus ($E_h=M_h$ and $\bd{p}_h=0$) these 
expressions can be rewritten as
\begin{equation}
\omega+M_h=E^\prime_h ; \hspace*{1cm} \bd{q} = \bd{p}^\prime_h
\end{equation}
and the invariant hadronic mass ($W$) takes the form
\begin{equation}
W^2=E^{\prime 2}_h - \bd{p}^{\prime 2}_h = (\omega+M_h)^2 - \bd{q}^2
\label{eq4}
\end{equation}

From eq.(\ref{eq2}) we obtain that
\begin{equation}
Q^2=-q_\mu^2=\bd{q}^2-\omega^2=4 E_0 E^\prime_0 \sin^2\frac{\theta_e}{2}
\label{eq5}
\end{equation}
where ultrarelativistic electrons have been assumed (the electron mass is 
neglected).

In the case of {\it elastic} scattering the invariant hadronic mass
coincides with the mass of the target ($W=M_h$), and eq.(\ref{eq4}) leads 
to
\begin{equation}
\omega=(M_ h^2+q^2)^{1/2}-M_h=E^\prime_h - M_h =T_h
\end{equation}
that is the kinetic energy of the final nucleus:
\begin{equation}
\omega=T_h=\frac{Q^2}{2M_h}
\end{equation}

Assuming that the two--neutron halo nucleus has no excited bound 
states (as it happens in $^6$He) an increase of
the energy transfer to the nucleus will break the halo system into its 
three constituents. The energy and momentum of the final hadronic system
can then be written as 
\begin{eqnarray}
E^\prime_h &=& E^\prime_1+E^\prime_2+E^\prime_3 \nonumber \\
\bd{p}^\prime_h&=&\bd{p}^\prime_1+\bd{p}^\prime_2+\bd{p}^\prime_3
\label{eq7}
\end{eqnarray}
where $(E_i,\bd{p}^\prime_i)$ ($i=1,2,3$) are the energy and momentum
of the three fragments. eqs. (\ref{eq1}) to (\ref{eq4}) are then valid
in this region after substitution on them of eq. (\ref{eq7}). Processes
involving excitations and one nucleon knockout from the core may take 
place at higher energies, and will not be discussed here.

\begin{figure}[ht]
\centerline{\psfig{figure=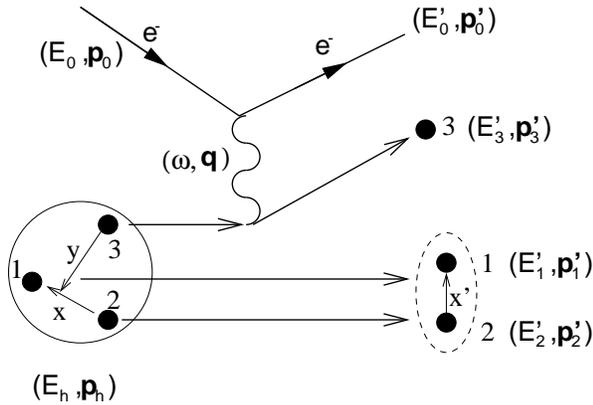,width=8.5cm,%
bbllx=6.1cm,bblly=6.5cm,bburx=15.2cm,bbury=21.2cm,angle=270}}
\vspace{0.2cm}
\caption[]{
Scheme of the reaction and definition of the coordinates
used to describe the exclusive $(e,e^\prime x)$ process on a two--neutron
halo nucleus. Only one of the constituents of the halo nucleus is considered
to interact with the electron.
}
\end{figure}

These breakup processes will be studied assuming that the 
whole energy and momentum transfer are absorbed by one of the 
constituents (participant), which is knocked out from the halo 
system. The other two constituents 
are mere spectators. The large extension of the
nucleus supports this picture, since the probability for
simultaneous interaction with the electron of two of the nuclear
constituents is small. The scheme of the reaction can be
seen in fig.2. If we consider constituent 3 as the participant
one, we then have
\begin{eqnarray}
E^\prime_3 &=& E_3 + \omega  \nonumber \\
\bd{p}^\prime_3 &=& \bd{p}_3 + \bd{q} \nonumber \\
E_h-E_3 &=& E^\prime_1 + E^\prime_2 \nonumber \\
\bd{p}_h - \bd{p}_3&=&\bd{p}^\prime_1 + \bd{p}^\prime_2 
\label{eq6}
\end{eqnarray} 
where $(E_3,\bd{p}_3)$ are the energy and momentum of the participant
constituent inside the halo nucleus.

In the target frame ($E_h=M_h$ and $\bd{p}_h=0$) we have that
$\bd{p}_3=-(\bd{p}^\prime_1 + \bd{p}^\prime_2)$, and
\begin{equation}
\omega+M_h=E^\prime_h=m_1+m_2+m_3+T_1+T_2+T_3.
\end{equation}
Assuming non relativistic kinetic energies for the three fragments
we have
\begin{eqnarray}
\lefteqn{
\omega-B_h=\frac{p^{\prime 2}_1}{2 m_1}+\frac{p^{\prime 2}_2}{2m_2}
+\frac{p^{\prime 2}_3}{2m_3}=} \nonumber \\ &&
E^\prime_{x}+\frac{p_3^2}{2(m_1+m_2)}
+\frac{(\bd{p}_3+\bd{q})^2}{2m_3}
\label{eq10}
\end{eqnarray}
where $B_h$ is the halo binding energy, $m_i$ and $T_i$ are the masses 
and final kinetic energies of the three constituents of the halo nucleus, 
and $E^\prime_{x}$ is the kinetic energy 
of the system made by particles 1 and 2 referred to its own center of mass.

From eq.(\ref{eq10}) one can see that
for a fixed value of the momentum transfer $\bd{q}$ the variation in the
energy transfer $\omega$ is connected to the different values of the internal
momentum $\bd{p}_3$ of the participant particle. Actually, for a fixed
value of the momentum transfer, electron--nucleus scattering cross sections show
a broad peak known as the quasielastic peak whose center is placed at $\bd{p}_3=0$.
Neglecting the binding energy $B_h$ and the internal energy $E^\prime_{x}$
we obtain from eq.(\ref{eq10}) the following value for the energy transfer
in the center of the quasielastic peak
\begin{equation}
\omega_{\mbox{\scriptsize q.p.}} \approx\frac{q^2}{2 m_3}
\label{qep}
\end{equation}
that corresponds to the kinetic energy transferred to a particle
of mass $m_3$ at rest.
Reactions taking place in a quasielastic peak region are interpreted 
as processes where a single constituent in 
the nucleus absorbs the whole energy and momentum transfer, being 
ejected from the nucleus.

When the participant particle is the core, the center of the
quasielastic peak is at smaller energy than when the
participant particle is one of the halo neutrons. In the 
$^6$He case $\omega_{\mbox{\scriptsize q.p.}}$ is approximately four
times smaller for a participant $^4$He than for a participant neutron.

In this paper the exclusive $(e,e^\prime x)$ reactions will be
investigated in the so called {\it perpendicular kinematics}, where
the energy and momentum transfer are maintained fixed. In this
kinematics, for a given electron beam with energy and momentum
$(E_0,\bd{p}_0)$, the scattering angle $\theta_e$ and the energy of 
the outgoing electron $E^\prime_0$ are fixed during the experiment.
From (\ref{eq2}) and (\ref{eq5}) it is clear that $\theta_e$ and 
$E^\prime_0$ can be chosen in such a way that $\omega$ and $\bd{q}$ take 
the desired values. In particular, for a fixed value of $\bd{q}$ the 
energy transfer will be taken as shown in (\ref{qep}). We are then
working in the center of a quasielastic peak.

The energy transfer in the center of the quasielastic peak obviously
increases with the momentum transfer $q$. If this momentum 
is such that  $\omega$ is larger than the
separation energy of the nucleons in core, then a
neutron detected in coincidence with the electron could come from
the core instead of coming from the halo. Therefore if  
$\omega$ is maintained below the separation
energy of the nucleons in the core we guarantee that
the halo nucleus is broken in only its three constituents. In 
particular, for $^6$He, since
the neutron or proton separation energy in $^4$He is around 20 MeV
a momentum transfer smaller than 200 MeV/c would satisfy this condition.
At the same time a value of $\omega$ below
20 MeV is also unable to excite the $\alpha$ particle to its first
excited state.

From eq.(\ref{eq6}) and assuming that $E_3^2=\bd{p}_3^2+m_3^2$ we
can write
\begin{equation}
\left( p_3^2+q^2+2p_3q\cos(\theta_3)+m_3^2\right)^{1/2} =
\left( p_3^2+m_3^2\right)^{1/2} + \omega
\label{balance}
\end{equation}
If we now take $q=200$ MeV/c and let the energy transfer
take the value  $q^2/2m_\alpha$ ($\omega \approx 5$ MeV), it is then
simple to see from eq.(\ref{balance}) that this energy can be absorbed
by a participant $\alpha$ particle ($m_3=m_\alpha$) with rather low 
internal momentum
(smaller than a few MeV/c). If for the same value of $\omega$ we
consider that the participant particle is one of the halo neutrons
($m_3=m_n$)
the required internal momentum of the neutron has to be very large
(typically of several hundreds of MeV/c). However the probability
of finding a halo neutron in $^6$He with such a large momentum is
basically zero \cite{kor94,gar97b}.
In the same way, from eq.(\ref{balance}) one sees that 
if $\omega=q^2/2m_n$
($\omega \approx 20$ MeV for $q=200$ MeV/c) a participant neutron
with small momentum can absorb such an energy, but it can not be
absorbed by a participant $\alpha$ particle.

Therefore, assuming that the reaction picture shown in
fig.2 is valid (a single constituent acts as participant while the
other two are just spectators), we can then conclude that an
energy transfer in the center of the $\alpha$ quasielastic peak
($\omega=q^2/2m_\alpha)$ 
selects the $\alpha$ as participant particle, and a halo neutron 
if $\omega$ is in the center of the neutron quasielastic peak
($\omega=q^2/2m_n)$.

\section{Elastic scattering}

The general expression for the elastic electron scattering cross
section can be found for instance in \cite{for66}. In particular
for a nucleus with spin zero, as $^6$He, the elastic cross section
in the nucleus frame  takes the form
\begin{equation}
\frac{d\sigma}{d\Omega_e}=\frac{Z^2 \sigma_M}{f_{\scriptsize rec}} 
       \frac{Q^4}{q^4} |F_{\scriptsize ch}(q) |^2
\label{ee}
\end{equation}
where the ultrarelativistic limit for the electrons has been assumed, 
$\Omega_e$ defines the direction of the outgoing electron, and
\begin{equation}
\sigma_M= \frac{\alpha^2 \cos^2{(\theta_e/2)}}{4 E_0^2 \sin^4{(\theta_e/2)}};
\hspace*{0.3cm}
f_{\scriptsize rec} = 1+\frac{2 E_0 \sin{(\theta_e/2)}}{M_h}
\label{smott}
\end{equation}
are the Mott cross section and the recoil factor.

The charge form factor $F_{\scriptsize ch}(q)$ is given by
\begin{equation}
F_{\scriptsize ch}(q)= \frac{4\pi}{Z} 
   \int \frac{\sin (qr)}{qr} \rho(r) r^2 dr
\label{eq14}
\end{equation}
where $\rho(r)$ is the charge density.

Assuming that the picture of a core plus several halo neutrons is valid
to describe neutron halo nuclei, the charge density should be 
practically identical to the one of the core. Therefore, since the neutron
contribution to the charge form factor will be negligible at moderate
$q$--values, the charge form factor should be
the one obtained in an elastic electron--core scattering process.

\begin{figure}[ht]
\centerline{\psfig{figure=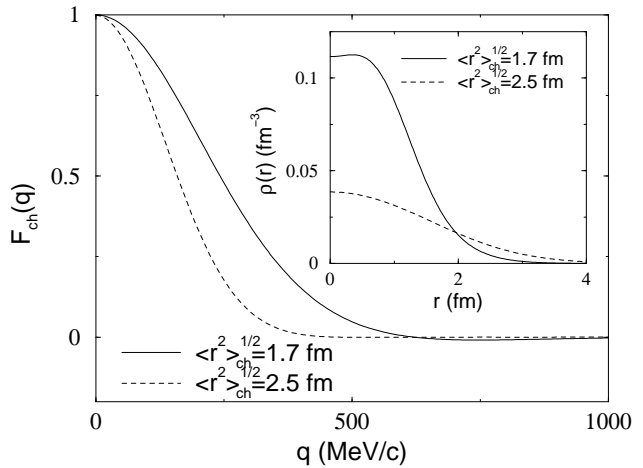,width=5cm,%
bbllx=4.0cm,bblly=6.5cm,bburx=15.2cm,bbury=20.2cm,angle=270}}
\vspace{2cm}
\caption[]{
Charge form factor after elastic electron scattering from $^6$He
assuming that the charge density corresponds to the $\alpha$ charge density
(solid line), and assuming that charge is spread out over the whole $^6$He
nucleus (dashed line). The inset shows the charge density used in both cases.
}
\end{figure}

To simplify the analysis we take the experimental
charge density of $^4$He that can be found in table V
of \cite{vri87}. The root mean square radius of the charge distribution
is 1.68 fm, and the charge form factor is shown in fig.3 by the
solid line. This should then be the charge form factor obtained
after elastic electron scattering on $^6$He provided that the
three--body picture is correct. 

The other limit would be to consider that the charge is not contained
only in the core but distributed like matter over the whole nucleus.
In that case the charge form factor would be the one shown in fig.3
by the dashed line. The form factor is now narrower due to the larger
value of the charge radius. This curve has been obtained by 
modifying the range
of the gaussians describing the $^4$He charge density in \cite{vri87}
in order to get a  root mean square radius of the charge distribution
of 2.5 fm, similar to the size of $^6$He.

From fig.3 we can conclude that elastic electron scattering from $^6$He,
or in general from neutron halo nuclei, is an excellent tool to determine
to what extent the few--body description is valid. The inset in fig.3 shows
the charge density for two cases mentioned above.

Center of mass corrections, also contained in the recoil factor of
eq.(\ref{ee}), may slightly modify the form factor and the charge r.m.s.
radius of $^6$He relative to that of $^4$He. These effects could be
accounted for using the three--body wave functions employed in the next
sections. However, to give a first  estimate of the size of the 
differential cross sections the two extreme pictures presented in
fig.3 serve the purpose. For instance, for $E_0 \approx 200$ MeV and
$\theta_e\approx 30$ degrees, the differential cross section is of
around 100 $\mu b$.

\section{Quasielastic scattering}

In this section we enter in the study of the coincidence $(e,e^\prime x)$
processes, with $x$ either a halo neutron or the core. The scheme 
of the reaction was shown in fig.2. 
A three--body system with energy and momentum $(E_h,\bd{p}_h)$ 
interacts with an electron with energy and momentum $(E_0,\bd{p}_0)$.
The whole energy and momentum transfer $(\omega,\bd{q})$ is absorbed
by one of the three constituents of the nucleus (constituent 3) that
is ejected out from the three-body halo system. The energies and momenta
in the final state are denoted by $(E^\prime_i,\bd{p}^\prime_i)$ where
$i=0$ for the electron and $i=1,2,3$ for the three halo constituents.

Together with the impulse approximation (the virtual photon is absorbed 
by a single constituent) we also neglect the Coulomb distortion of the
ingoing and outgoing electrons. They are then described as plane waves. 
At the same time the distortion due to the final strong
interaction between particle 3 (participant) and the two--body subsystem 
made by particles 1 and 2 (spectators) is also neglected, and the 
knocked out constituent 
is also described as a plane wave. On the contrary, the interaction
between particles 1 and 2 is included in the calculation. This 
interaction is what in refs. \cite{gar96,gar97} is referred to as final 
state interaction, and plays an essential role in the description
of the halo nucleus fragmentation reactions. Our picture is then
similar to that described for instance in \cite{fru84,bof93}
for $(e,e^\prime N)$ reactions in PWIA. The only difference (but 
important) is that our {\it residual} nucleus is an unbound two--body
system, while they consider that the residual nucleus is in a 
well defined bound state. 

In appendix A it is shown that working in the frame of the three--body
halo nucleus ($\bd{p}_h=\bd{p}_1+\bd{p}_2+\bd{p}_3=0$) the transition 
matrix element of the reaction has the form (\ref{matrix})

\begin{eqnarray}
S_{fi}&=&\frac{i}{V^3} \frac{e^2}{q_\rho^2} (2 \pi)^{17/2}
\delta(E_i-E_f) \delta^3(\bd{P}_i-\bd{P}_f)  \\ && \hspace*{-1cm}
\bar{u}(\bd{p}^\prime_0,\sigma^\prime_0) \gamma_\mu
u(\bd{p}_0,\sigma_0) 
\sum_{\sigma_3} J^\mu_{\sigma_3^\prime,\sigma_3}(\bd{q},\bd{p}_3)
M^{JM}_{s_{12} \sigma_{12}, s_3 \sigma_3}(\bd{p}^\prime_x,\bd{p}^\prime_y)
\nonumber
\end{eqnarray}
where $E_i=E_0+E_h$ and $E_f=E^\prime_0+E^\prime_1+E^\prime_2+E^\prime_3$
are the initial and final total energies, and analogously for the
initial and final total momenta $\bd{P}_i$ and $\bd{P}_f$. The Jacobi
coordinates $\bf{x}$ and $\bf{y}$ (see fig.2) are defined in (\ref{jac}) and
their conjugated momenta $\bd{p}_x$ and $\bd{p}_y$ are given in
(\ref{mom}). The momenta in the final state are denoted with primes.
The spinor  $u(\bd{p},\sigma)$ describes a free electron
with momentum $\bd{p}$ and third spin component $\sigma$. 
$J^\mu_{\sigma_3^\prime,\sigma_3}(\bd{q},\bd{p}_3)$ is the matrix element 
in momentum
space of the current operator connecting states of the particle 3 with
spin projections $\sigma_3$ and $\sigma^\prime_3$ (eq.(\ref{cur})).
In (eq.(\ref{sudd})) we have also defined 
$M^{JM}_{s_{12} \sigma_{12}, s_3 \sigma^\prime_3}(\bd{p}^\prime_x,\bd{p}^\prime_y)$,
which is the overlap between the initial three--body halo wave 
function ($\Psi^{JM}(\bd{x},\bd{y})$) and the continuum wave function 
of the final two--body subsystem 
($w^{s_{12},\sigma_{12}}(\bd{x},\bd{p^\prime_x})$).
When
the interaction between particles 1 and 2 is neglected 
$M^{JM}_{s_{12} \sigma_{12}, s_3 \sigma^\prime_3}(\bd{p}^\prime_x,\bd{p}^\prime_y)$
becomes the Fourier transform (normalized to 1) of the
three-body halo wave function. $s_{12}$ and $\sigma_{12}$ are the total
spin and its third component of the two--body system made by particles
1 and 2.

We can now compute the differential cross section as the transition 
probability per unit volume and unit time ($|S_{fi}|^2/VT$) divided
by the flux of incident particles ($p_0/V E_0$ in the halo nucleus 
frame), divided by the number of target particles per unit volume
($1/V$) and multiplied by the number of final states that is
given by
\begin{eqnarray}
\lefteqn{
V \frac{d\bd{p}^\prime_0}{(2\pi)^3}
V \frac{d\bd{p}^\prime_1}{(2\pi)^3}
V \frac{d\bd{p}^\prime_2}{(2\pi)^3}
V \frac{d\bd{p}^\prime_3}{(2\pi)^3} =} \nonumber \\ &&
V \frac{d\bd{p}^\prime_0}{(2\pi)^3}
V \frac{d\bd{p}^\prime_3}{(2\pi)^3}
V \frac{d\bd{p}^\prime_x}{(2\pi)^3}
V \frac{d\bd{p}^\prime_y}{(2\pi)^3}
\label{fst}
\end{eqnarray}

After averaging over initial states and summing over final states 
we obtain
\begin{eqnarray}
\frac{d^{12}\sigma}
   {d\bd{p}^\prime_0 d\bd{p}^\prime_3 d\bd{p}^\prime_x d\bd{p}^\prime_y} &=&
(2\pi) \frac{e^4}{q_\rho^4} \frac{E_0}{p_0} \delta(E_i-E_f)
\delta(\bd{P}_i-\bd{P}_f)  \nonumber \\ &&  
\frac{1}{2} \sum_{\sigma_0, \sigma^\prime_0}
|\bar{u}(\bd{p}^\prime_0, \sigma^\prime_0) \gamma_\mu u(\bd{p}_0, \sigma_0)|^2
 \label{cross}   
\\  & & \hspace*{-3.cm} 
\frac{1}{2J+1} \sum_{M \sigma_3} \sum_{s_{12} \sigma_{12}}
\left| \sum_{\sigma^\prime_3} J_{\sigma_3,\sigma^\prime_3}^\mu(\bd{q},\bd{p}_3)
M^{JM}_{s_{12} \sigma_{12}, s_3 \sigma_3^\prime}
(\bd{p}^\prime_x,\bd{p}^\prime_y) \right|^2 \nonumber
\end{eqnarray}
In this expression the integral over $\bd{p}^\prime_y$ can be immediately done
by use of the delta of momenta, getting that 
$\bd{p}^\prime_y=\bd{q}-\bd{p}^\prime_3=-\bd{p}_3$ (see eq.(\ref{relat})).

In the ultrarelativistic limit it is known that \cite{don75}
\begin{equation}
\frac{1}{2} \sum_{\sigma_0, \sigma^\prime_0}
|\bar{u}(\bd{p}^\prime_0, \sigma^\prime_0) \gamma_\mu u(\bd{p}_0, \sigma_0)|^2=
\frac{1}{2E_0E^\prime_0} \eta_{\mu \nu}
\label{lept}
\end{equation}
where the leptonic tensor $\eta_{\mu\nu}$ is
\begin{equation}
\eta_{\mu \nu}=p_{0\mu} p^\prime_{0\nu}+p_{0\nu} p^\prime_{0\mu}+\frac{1}{2}
q_\rho^2 g_{\mu \nu}
\end{equation}
and $g_{\mu \nu}$ is the diagonal metric $(1,-1,-1,-1)$.

For a particle 3 with spin 0 or 1/2  the matrix
\begin{equation}
M^{* JM}_{s_{12} \sigma_{12}, s_3 \sigma_3}
(\bd{p}^\prime_x,\bd{p}_3) 
M^{JM}_{s_{12} \sigma_{12}, s_3 \sigma_3^\prime}
(\bd{p}^\prime_x,\bd{p}_3)
\end{equation}
is diagonal and has both diagonal terms equal. The differential 
cross section can then be written
\begin{eqnarray}
\frac{d^9\sigma}
   {d\bd{p}^\prime_0 d\bd{p}^\prime_3 d\bd{p}^\prime_x} & = &
(2 \pi)^3 \frac{2 \alpha^2}{q_\rho^4} \frac{1}{E_0 E^\prime_0}
\delta(E_i-E_f) \eta_{\mu\nu}R^{\mu\nu} \nonumber \\ && \hspace*{-0.5cm} 
\frac{1}{2J+1} \sum_{M \sigma_3 s_{12} \sigma_{12}}
\left|
M^{JM}_{s_{12} \sigma_{12}, s_3 \sigma_3}(\bd{p}^\prime_x,\bd{p}_3)
                 \right|^2
\label{eq21}
\end{eqnarray}
with $\alpha=e^2/4\pi$, and where we have defined the hadronic tensor
\begin{equation}
R^{\mu \nu} = \frac{1}{2s_3+1} \sum_{\sigma_3 \sigma^\prime_3}
|J^\mu_{\sigma_3 \sigma_3^\prime}(\bd{q},\bd{p}_3)|^2
\end{equation}
that is the hadronic tensor associated to an electron--particle 3 
scattering process. 

When particle 3 is one of the halo neutrons the electron-neutron cross 
section enters in (\ref{eq21}), and has the standard form 
\begin{eqnarray}
\sigma^{en}(\bd{q},\bd{p}_3)=
\frac{2 \alpha^2}{q_\rho^4} \frac{E^\prime_0}{E_0}
\eta_{\mu\nu}R^{\mu\nu} & = & \nonumber \\ && \hspace*{-4cm}
   \sigma_M \left(
   V_L {\cal R}^{(n)}_L+V_T{\cal R}^{(n)}_T+
   V_{LT}{\cal R}^{(n)}_{LT}+V_{TT}{\cal R}^{(n)}_{TT} 
                                              \right)
\label{nucleon}
\end{eqnarray}
where the kinematic factors V's and the response functions 
${\cal R}$'s are given by
\begin{equation}
\begin{array}{ll}
 V_L = \frac{Q^4}{q^4} & {\cal R}^{(n)}_L = R^{00} \\
 V_T = \tan^2(\frac{\theta_e}{2}) + \frac{1}{2} \frac{Q^2}{q^2} 
                     & {\cal R}^{(n)}_T = R^{11} + R^{22} \\
 V_{LT} =  \frac{1}{\sqrt{2}} \frac{Q^2}{q^2} 
  \left(  \tan^2(\frac{\theta_e}{2}) + \frac{Q^2}{q^2} \right)^{1/2}
           & {\cal R}^{(n)}_{LT} = -2 \sqrt{2} R^{10} \\
 V_{TT} =  -\frac{1}{2} \frac{Q^2}{q^2} 
           & {\cal R}^{(n)}_{TT} = R^{22} - R^{11}\\
\end{array}
\label{vr}
\end{equation}
as it can be found for instance in  \cite{for83,ras89}.

If particle 3 is the $\alpha$--particle the electron--$\alpha$ cross
section reduces to 
\begin{equation}
\sigma^{e\alpha}(\bd{q},\bd{p}_3)=\frac{2 \alpha^2}{q_\rho^4} 
\frac{E^\prime_0}{E_0}\eta_{\mu\nu}R^{\mu\nu}=
\sigma_M V_L {\cal R}^{(\alpha)}_L
\label{alpha}
\end{equation}

Explicit expressions for ${\cal R}^{(n)}$ and ${\cal R}^{(\alpha)}$ are
given in the next section.

Taking this into account the nine-differential cross section (\ref{eq21})
takes now the form
\begin{eqnarray}
\frac{d^9\sigma}
   {d\bd{p}^\prime_0 d\bd{p}^\prime_3 d\bd{p}^\prime_x} & = &
(2 \pi)^3  \frac{1}{E^{\prime 2}_0}
\delta(E_i-E_f) \sigma^{e3}(\bd{q},\bd{p}_3) \nonumber \\ && \hspace*{-1cm} 
\frac{1}{2J+1} \sum_{M \sigma_3 s_{12} \sigma_{12}}
\left|
M^{JM}_{s_{12} \sigma_{12}, s_3 \sigma_3}(\bd{p}^\prime_x,\bd{p}_3)
                 \right|^2
\label{dif9}
\end{eqnarray}
where the delta of energies imposes that 
\begin{eqnarray}
E_i-E_f&=&M_h+\omega-(p^{\prime 2}_3 + m_3^2)^{1/2} - E^\prime_x - 
\nonumber \\ &&
    \frac{(\bd{q}-\bd{p}^\prime_3)^2}{2(m_1+m_2)} -(m_1+m_2)=0,
\label{delta}
\end{eqnarray}
$M_h$ is the mass of the halo nucleus and
\begin{equation}
E^\prime_x=\frac{p^{\prime 2}_x}{2 \mu_{12}}
\label{inter}
\end{equation}
is the internal energy of the final two--body system.

Making use of the delta function one can integrate over $p^\prime_3$, that 
leads to 
\begin{eqnarray}
\frac{d^8\sigma}{dE^\prime_0 d\Omega_{p^\prime_0} d\Omega_{p^\prime_3}
d\bd{p}^\prime_x} &=& (2\pi)^3 \frac{p^\prime_3 E^\prime_3 (m_1+m_2)}{M_h}
f_{rec} \sigma^{e3}(\bd{q},\bd{p}_3)  \nonumber \\ && \hspace*{-2cm}
\frac{1}{2J+1} \sum_{M \sigma_3 s_{12} \sigma_{12}}
\left|
M^{JM}_{s_{12} \sigma_{12}, s_3 \sigma_3}(\bd{p}^\prime_x,\bd{p}_3)
                 \right|^2
\end{eqnarray}
where $\bd{p}_3=\bd{p}^\prime_3-\bd{q}$, $p^\prime_3$ is obtained from 
(\ref{delta}),  the angles $\Omega_{p^\prime_0}$ and $\Omega_{p^\prime_3}$ 
define the directions of $\bd{p}\prime_0$ and $\bd{p}^\prime_3$, respectively,
 and the recoil factor $f_{rec}$ is given by
\begin{equation}
f_{rec}=\left( 
1+\frac{\omega p^\prime_3-
E^\prime_3 q \cos\theta_{p^\prime_3}}{M_h p^\prime_3}
                                                \right)^{-1}
\end{equation}

Finally, from  (\ref{inter}) we have that
$d\bd{p}^\prime_x=p^\prime_x \mu_{12} dE^\prime_x d\Omega_{p^\prime_x}$,
and then one gets the familiar expression
\begin{eqnarray}
\lefteqn{
\frac{d^6\sigma}
   {dE^\prime_0 d\Omega_{p^\prime_0} d\Omega_{p^\prime_3} dE^\prime_x} =}
 \nonumber \\ && 
(2 \pi)^3  \frac{p^\prime_3 E^\prime_3 (m_1+m_2)}{M_h} f_{rec}
\sigma^{e3}(\bd{q}, \bd{p}_3)  S(E^\prime_x,\bd{p}_3)
\label{factor}
\end{eqnarray}
where 
\begin{eqnarray}
\lefteqn{
S(E^\prime_x,\bd{p}_3) =} \nonumber \\ &&
\frac{p^\prime_x \mu_{12}}{2J+1}
\sum_{M \sigma_3 s_{12} \sigma_{12}} \int d\Omega_{p^\prime_x}
\left|
M^{JM}_{s_{12} \sigma_{12}, s_3 \sigma_3}(\bd{p}^\prime_x,\bd{p}_3)
                 \right|^2
\label{spec}
\end{eqnarray}

We have then  arrived in eq.(\ref{factor}) to the usual factorization
of the cross section for an $(e,e^\prime N)$ process in PWIA 
\cite{fru84,don75,for83,ras89}. The differential cross section is written
as product of the cross section of an electron--particle 3 scattering
process and a function $S(E^\prime_x,\bd{p}_3)$ called Spectral
Function. The spectral function is interpreted as the probability 
for the electron to remove a particle from the nucleus with internal
momentum $\bd{p}_3$ leaving the residual system with internal energy 
$E^\prime_x$ \cite{fru84}. 

In stable nuclei the residual nucleus is usually considered to 
be in a bound state, in such a way that its internal energy can only 
take discrete values corresponding to the different
excited states of the residual nucleus. If the missing energy 
(defined as the part
of the initial energy that is not transformed into kinetic energy)
is kept within the appropriate limits it is possible to select a
specific state for the residual nucleus. This amounts to fix the
internal energy $E^\prime_x$ to the value corresponding to a given
excited state. The spectral function then gives the momentum distribution
of the particle such that after removal leaves the residual nucleus
in such an excited state.

In our case the residual nucleus is not bound, but it is made by two particles
flying together in the continuum. In other words, the internal energy 
$E^\prime_x=p^{\prime 2}_x/(2\mu_{12})$ takes continuum values, and it
is not possible to select a definite state for the final two--body system
made by particles 1 and 2. Therefore integration over $E^\prime_x$ is required,
and 
\begin{equation} 
\frac{d^5\sigma}
   {dE^\prime_0 d\Omega_{p^\prime_0} d\Omega_{p^\prime_3}} =
\int dE^\prime_x 
\frac{d^6\sigma}
   {dE^\prime_0 d\Omega_{p^\prime_0} d\Omega_{p^\prime_3} dE^\prime_x}
\label{dif5}
\end{equation}

Note that the cross section $\sigma^{e3}$ and the recoil factor 
depend on $\bd{p}^\prime_3$
($\bd{p}_3=\bd{p}^\prime_3-\bd{q}$), and  $p^\prime_3$ is obtained
from (\ref{delta}) that depends on $E^\prime_x$. Therefore, the integration
in (\ref{dif5}) involves the whole expression in (\ref{factor}), and the
factorization disappears.

Nevertheless, when the two--body system has resonances at low energy, 
$E^\prime_x$ can be neglected in eq.(\ref{delta}), and the value of 
$p^\prime_3$, and therefore $\sigma^{e3}$ and the recoil factor, are to 
a good approximation independent of
$E^\prime_x$. Now the integral over $E^\prime_x$ involves only the spectral
function, and we can then write
\begin{equation}
\frac{d^5\sigma}
   {dE^\prime_0 d\Omega_{p^\prime_0} d\Omega_{p^\prime_3}} =
(2 \pi)^3  \frac{p^\prime_3 E^\prime_3 (m_1+m_2)}{M_h} f_{rec}
\sigma^{e3}(\bd{q}, \bd{p}_3)  n(\bd{p}_3)
\label{noep}
\end{equation}
where
\begin{equation}
n(\bd{p}_3)=\int dE^\prime_x S(E^\prime_x,\bd{p}_3)
\label{eq35}
\end{equation}
that can be interpreted as the momentum distribution of the
constituent particle 3.

If we now introduce eq.(\ref{nucleon}) or (\ref{alpha}), we can finally write
\begin{eqnarray}
\lefteqn{
\left(
\frac{d^5\sigma}
   {dE^\prime_0 d\Omega_{p^\prime_0} d\Omega_{p^\prime_3}} 
\right)^{(e,e^\prime n)}=
(2 \pi)^3  \frac{p^\prime_3 E^\prime_3 (m_1+m_2)}{M_h} f_{rec} } 
\nonumber \\ &&
\sigma_M (V_L W^{(n)}_L + V_T W^{(n)}_T +
        V_{LT} W^{(n)}_{LT} + V_{TT} W^{(n)}_{TT})
\label{fincross}
\end{eqnarray}
or
\begin{eqnarray}
\left(
\frac{d^5\sigma}
   {dE^\prime_0 d\Omega_{p^\prime_0} d\Omega_{p^\prime_3}} 
\right)^{(e,e^\prime \alpha)}&=&
(2 \pi)^3  \frac{p^\prime_3 E^\prime_3 (m_1+m_2)}{M_h} f_{rec}  
\nonumber \\ &&
\sigma_M V_L W^{(\alpha)}_L
\label{across}
\end{eqnarray}
for the cases of participant neutron and participant $\alpha$, respectively.
The $W$'s are the usual structure functions, and they are product
of the momentum distribution $n(\bd{p}_3)$ and the corresponding
response function ${\cal R}^{(3)}$. The five--differential
cross section (\ref{fincross}) is formally identical to
the expressions given for instance in \cite{fru84,for83,ras89} for the
$(e,e^\prime N)$ cross section in PWIA assuming a residual nucleus in a
bound state. The response function for $(e,e^\prime \alpha)$ will be
specified later.

Up to now we have always considered that the particle detected in coincidence
with the electron is the participant particle (particle 3). However the case 
where one of the spectator particles (particle 1 or 2) is detected in 
coincidence with the electron is also of interest. When one of the neutrons
in the two--neutron halo nucleus is knocked out by the electron we actually
have two neutrons in the final state, one of them participant and the 
other one spectator. When one neutron is detected it is obviously not possible 
to know which one was detected, and therefore both cases should be considered.

In eq.(\ref{cross}) we can use the delta of momenta to integrate over 
$\bd{p}^\prime_3$ instead of $\bd{p}^\prime_y$ 
($\bd{p}^\prime_3=\bd{q}-\bd{p}^\prime_y$). Following the same steps we
then arrive to an expression analogous to (\ref{dif9})
\begin{eqnarray}
\frac{d^9\sigma}
   {d\bd{p}^\prime_0 d\bd{p}^\prime_x d\bd{p}^\prime_y} &=&
(2 \pi)^3  \frac{1}{E^{\prime 2}_0}
\delta(E_i-E_f) \sigma^{e3}(\bd{q},\bd{p}^\prime_y) \nonumber \\ &&
\hspace*{-1cm}
\frac{1}{2J+1} \sum_{M \sigma_3 s_{12} \sigma_{12}}
\left|
M^{JM}_{s_{12} \sigma_{12}, s_3 \sigma_3}(\bd{p}^\prime_x,\bd{p}^\prime_y)
                 \right|^2
\end{eqnarray}

From eq.(\ref{mom}) we obtain
\begin{eqnarray}
\bd{p}^\prime_1& = & \bd{p}^\prime_x + \frac{m_1}{m_1+m_2} \bd{p}^\prime_y
                                \nonumber \\
\bd{p}^\prime_2& = & -\bd{p}^\prime_x + \frac{m_2}{m_1+m_2} \bd{p}^\prime_y
\label{surv}
\end{eqnarray}
from which it is easy to see that $d\bd{p}^\prime_x d\bd{p}^\prime_y=
d\bd{p}^\prime_i d\bd{p}^\prime_y$ ($i=1,2$).

Using now the delta of energies we can integrate over $p^\prime_y$ and then
get
\begin{eqnarray}
\frac{d^8\sigma}{d\bd{p}^\prime_0  d\bd{p}^\prime_i d\Omega_{p^\prime_y}} 
& = & (2\pi)^3 \frac{p^{\prime 2}_y}{E^\prime_0} f_{rec} 
\sigma^{e3}(\bd{q},\bd{p}^\prime_y) \nonumber \\ && \hspace*{-1cm} 
\frac{1}{2J+1} \sum_{M \sigma_3 s_{12} \sigma_{12}}
\left|
M^{JM}_{s_{12} \sigma_{12}, s_3 \sigma_3}(\bd{p}^\prime_x,\bd{p}^\prime_y)
                 \right|^2
\label{newc}
\end{eqnarray}
where $i=1,2$ and $p^\prime_y$ is obtained from (\ref{delta}) 
(using the relation $\bd{p}^\prime_y=\bd{q}-\bd{p}^\prime_3$ and
eq.(\ref{surv})).   The recoil
factor $f_{rec}$ is in this case the value of the inverse of the 
derivative of (\ref{delta}) with respect to $p^\prime_y$. 

From (\ref{newc}) we obtain the  five--differential cross section
\begin{equation}
\frac{d^5\sigma}{dE^\prime_0 d\Omega_{p^\prime_0} d\Omega_{p^\prime_i}}
= \int p^{\prime 2}_i dp^\prime_i d\Omega_{p^\prime_y} p^{\prime 2}_0
\frac{d^8\sigma}{d\bd{p}^\prime_0  d\bd{p}^\prime_i d\Omega_{p^\prime_y}}
\label{dif5b}
\end{equation}
that is analogous to (\ref{dif5}) and where the momentum $\bd{p}^\prime_i$
($i=1,2$) is now the momentum of one of the spectator particles in the
final state.

In principle the differential cross sections (\ref{dif5}) 
and (\ref{dif5b}) depend on five variables ($E^\prime_0$ and four angles).
However electron scattering experiments are normally performed in what is
known as {\it in plane} kinematics. This means that only particles in the 
scattering plane are detected. We then restrict ourselves to
processes in the $xz$--plane (see fig.1b), and the azimuthal angles 
$\varphi_{p^\prime_0}$, $\varphi_{p^\prime_3}$ (in (\ref{dif5})), and 
$\varphi_{p^\prime_i}$ (in (\ref{dif5b})) are taken equal to zero. 
The number of variables 
is reduced now to three. Working in perpendicular kinematics the 
energy transfer to the nucleus is kept fixed, and for a given incident 
electron energy $E_0$ one has $E^\prime_0=E_0-\omega$. There are then 
only two variables left, 
$\theta_{p^\prime_0}$ and $\theta_{p^\prime_3}$ in (\ref{dif5}), and
$\theta_{p^\prime_0}$ and $\theta_{p^\prime_i}$ in (\ref{dif5b}). 
Finally we also know that $\bd{p}^\prime_0=\bd{p}_0-\bd{q}$, and multiplying 
this expression by $\bd{p}^\prime_0$ we obtain in the ultrarelativistic limit
\begin{equation}
\cos \theta_{p^\prime_0}=\frac{E_0 \cos\theta_e - E^\prime_0}{q}
\end{equation}
and therefore the differential cross sections in the {\it in plane}
perpendicular kinematics are simply a function of the polar angle
$\theta_{p^\prime_3}$ in (\ref{dif5}) or $\theta_{p^\prime_i}$ in 
(\ref{dif5b}).

As mentioned in the previous section, 
in case that we neglect $E^\prime_x$ in (\ref{delta}) the expression
(\ref{noep}) is valid, and the observables can
be given as function either of $\theta_{p^\prime_3}$ or of the
internal momentum of the participant particle $p_3$. This is because
$\theta_{p^\prime_3}$ and $p_3$ are connected through the relation
\begin{equation}
p_3^2=(\bd{p}^\prime_3 - \bd{q})^2=
        p^{\prime 2}_3 + q^2-2qp^\prime_3 \cos\theta_{p^\prime_3}
\end{equation}
where $p^\prime_3$ is obtained from (\ref{delta}) provided that $E^\prime_x$
is neglected.

\section{Spectral function and response functions}
In order to obtain the coincidence differential cross sections 
given in (\ref{dif5}) and (\ref{dif5b}), we need to specify the
spectral function (\ref{spec}) of the halo nucleus and the
elementary electron--neutron responses, the electron--$\alpha$ responses,
and the cross sections.

In particular, for the spectral function we need the probability
function  
\begin{equation}
{\cal P}(\bd{p}^\prime_x,\bd{p}^\prime_y)=
\frac{1}{2J+1}
\sum_{M \sigma_3 s_{12} \sigma_{12}}
\left|
M^{JM}_{s_{12} \sigma_{12}, s_3 \sigma_3}(\bd{p}^\prime_x,\bd{p}^\prime_y)
                 \right|^2
\end{equation}

The calculation of this summation requires the knowledge of  
{\it i)} the intrinsic wave function of the three--body halo 
nucleus $\Psi^{JM}(\bd{x},\bd{y})$, {\it ii)} the continuum
wave function $w^{s_{12},\sigma_{12}}(\bd{x},\bd{p^\prime_x})$ of 
the spectator system in the final state, and {\it iii)} compute the 
overlap in eq.(\ref{sudd}).
When applying the model to a particular case the calculation of the wave
functions need in its turn the interactions between the three constituents
in the halo system. In the case of $^6$He, the neutron--neutron
and neutron--$\alpha$ interactions need to be specified.

\subsection{Electron--participant particle cross section}

When the participant particle is one of the neutrons from the halo the 
electron--neutron cross section is needed. Electron scattering by
a free nucleon is a well understood reaction, and the only uncertainties
come from the off--shell character of the nucleon interacting with the
electron. For halo neutrons off--shell effects in the nucleon current
can not be expected to be important, and we consider the CC1
prescription \cite{for83,ras89,cab93}. 

\begin{eqnarray}
\lefteqn{
J^\mu_{\sigma_3,\sigma^\prime_3}(\bd{p}_3,\bd{p}^\prime_3)= 
\bar{u}(\bd{p}^\prime_3, \sigma^\prime_3) } \\ &&
\left( 
(F_1(q^2)+F_2(q^2)) \gamma^\mu -
\frac{F_2(q^2)}{2M_{\mbox{\scriptsize nucleon}}}
(p_3^\mu + p^{\prime \mu}_s )
             \right)u(\bd{p}_3, \sigma_3) \nonumber
\end{eqnarray} 
subject to the condition of 
current conservation ($J^3_{\sigma_3,\sigma^\prime_3}=
J^0_{\sigma_3,\sigma^\prime_3} \omega/q$). 

With this prescription the expressions for the structure functions in
eq.(\ref{vr}) are given in appendix B.

If the participant particle is the core ($\alpha$--particle in our case)
we also need the $\sigma^{e\alpha}$ cross section. Since the $\alpha$--particle
has spin zero, only the component with $\mu=0$ is non-zero in (\ref{cur}), and
\begin{equation}
J^0(\bd{q})=\rho(\bd{q})=\frac{1}{(2\pi)^{3/2}}
\int d\bd{z} e^{i \bd{q} \cdot \bd{z}} \rho(\bd{z})
\end{equation} 
As a consequence only ${\cal R}_L$ in (\ref{vr}) is different from zero, and
then
\begin{equation}
\sigma^{e\alpha}= \sigma_M V_L {\cal R}_L = \sigma_M \frac{Q^4}{q^4}
\left| \rho(\bd{q}+\bd{p}_3) \right|^2
\end{equation} 

The charge density $\rho(\bd{r})$ for the $\alpha$--particle can be taken
from \cite{vri87}, where it is parameterized as a sum
of gaussians, as discussed in section III.

\subsection{Intrinsic wave functions}

The three-body wave function of the halo nucleus is obtained by solving
the Faddeev equations in coordinate space. The
total wave function $\Psi^{J M}$ of the three-body system (with total
spin $J$ and projection $M$) is written as a sum of three components,
each of them written as function of one of three possible sets of
Jacobi coordinates \cite{fed94}. For each Jacobi set we construct the 
hyperspherical coordinates ($\rho$, $\alpha$, $\Omega_x$, $\Omega_y$) 
defined in refs.\cite{zhu93,fed94}. The
volume element is given by $\rho^5 d\Omega d\rho$, where
$d\Omega=\sin^2\alpha \cos^2\alpha d\alpha d\Omega_x d\Omega_y$.
For each hyperradius $\rho$ $\Psi^{J M}$ is  expanded in a complete
set of generalized angular functions $\Phi^{(i)}_{n}(\rho,\Omega_i)$
\begin{equation}
\Psi^{J M}= \frac {1}{\rho^{5/2}}
  \sum_n f_n(\rho)
\sum_{i=1}^3 \Phi^{(i)}_{n}(\rho ,\Omega_i) \; ,
\label{tot}
\end{equation}
where $\rho^{-5/2}$ is related to the volume element, and the index $i$ 
refers to the three sets of Jacobi coordinates. 

The angular functions satisfy the angular part of the three Faddeev
equations:
\begin{eqnarray}
 {\hbar^2 \over 2m}\frac{1}{\rho^2}\hat\Lambda^2 \Phi^{(i)}_{n}
 +V_{jk} (\Phi^{(i)}_{n}+\Phi^{(j)}_{n J M}+
                \Phi^{(k)}_{n})& \equiv & \nonumber \\ && \hspace*{-3cm} 
{\hbar^2 \over 2m}\frac{1}{\rho^2} \lambda_n(\rho) \Phi^{(i)}_{n}  \; ,
\label{ang}
\end{eqnarray}
where $\{i,j,k\}$ is a cyclic permutation of $\{1,2,3\}$, $m$ is
an arbitrary normalization mass, $V_{jk}$ is the interaction between
particles $j$ and $k$, and $\hat\Lambda^2$ is the
$\rho$-independent part of the kinetic energy operator. The analytic
expressions for $\hat{\Lambda}^2$ and the kinetic energy operator can
for instance be found in \cite{fed94}.

The radial expansion coefficients $f_n(\rho)$ are obtained from
a coupled set of ``radial'' differential equations \cite{fed94}, i.e.
\begin{eqnarray}
   \left(-\frac{\rm d ^2}{\rm d \rho^2}
   -{2mE\over\hbar^2}+ \frac{1}{\rho^2}\left( \lambda_n(\rho) - Q_{n n}
+
  \frac{15}{4}\right) \right)f_n(\rho) &=&
 \nonumber \\ && \hspace*{-6cm}
  \sum_{n' \neq n}   \left(
   2P_{n n'}{\rm d \over\rm d \rho}
   +Q_{n n'}
   \right)f_{n'}(\rho)  \; , 
 \label{rad}
\end{eqnarray}
where the functions $P$ and $Q$ are defined as angular integrals:
\begin{equation}
   P_{n n'}(\rho)\equiv \sum_{i,j=1}^{3}
   \int d\Omega \Phi_n^{(i)\ast}(\rho,\Omega)
   {\partial\over\partial\rho}\Phi_{n'}^{(j)}(\rho,\Omega)  \; ,
\end{equation}
\begin{equation}
   Q_{n n'}(\rho)\equiv \sum_{i,j=1}^{3}
   \int d\Omega \Phi_n^{(i)\ast}(\rho,\Omega)
   {\partial^2\over\partial\rho^2}\Phi_{n'}^{(j)}(\rho,\Omega)  \; .
\end{equation}

After obtaining the three--body wave function (\ref{tot}) two of the 
components can be rotated to the third one, such that $\Psi^{J M}$ is 
written as function of a single set of Jacobi coordinates. In particular, 
to describe the reaction shown in fig.2 it is convenient to write
$\Psi^{J M}$ in terms of the Jacobi coordinates $\bd{x}$ and $\bd{y}$ 
shown in the same figure (note that the Jacobi coordinates are usually 
defined with some mass factors \cite{fed94} that for simplicity are not 
included in (\ref{jac})). 

The continuum wave function $w^{s_{12}, \sigma_{12}}(\bd{p}^\prime_x, \bd{x})$ 
describing the spectator two--body system in the final state is expanded 
in partial waves
\begin{eqnarray}
w^{s_{12}, \sigma_{12}}(\bd{p}^\prime_x, \bd{x})
& = & \label{2beq} \\ && \hspace*{-2cm}
\sqrt{\frac{2}{\pi}}
\frac{1}{p^\prime_x x} \sum_{j_{12} \ell_{12} m_{12}}
u_{\ell_{12} s_{12}}^{j_{12}}(p^\prime_x, x)
{\cal Y}_{j_{12} \ell_{12} s_{12}}^{m_{12}^*}(\Omega_{x}) \nonumber
\\ & & \hspace*{-2cm}
  \sum_{m_{\ell_{12}}=-\ell_{12}}^{\ell_{12}}
\langle \ell_{12} m_{\ell_{12}} ; s_{12} \sigma_{12} | j_{12} m_{12}
\rangle
i^{\ell_{12}} Y_{\ell_{12} m_{\ell_{12}}} (\Omega_{p^\prime_x})
\nonumber
\end{eqnarray}
where the radial functions $u_{\ell_{12} s_{12}}^{j_{12}}(p^\prime_x,
x)$ are obtained by solving the Schr\"{o}dinger equation with the
appropriate two--body potential \cite{gar97}. $\ell_{12}$ is the
relative orbital angular momentum between particles 1 and 2, that
coupled to $s_{12}$ gives the total angular momentum $j_{12}$. When the 
final state 
interaction between the spectator particles is neglected the expansion
(\ref{2beq}) becomes the usual expansion of a plane wave in terms of
spherical Bessel functions.

\subsection{Wave functions overlap and spectral function}

The cross sections derived in section IV are given in terms of the 
overlap between the initial two--neutron halo wave function and
the continuum wave function of the spectator system in the final
state (eq.(\ref{sudd})). After writing (\ref{tot}) in terms
of the Jacobi coordinates defined in (\ref{jac}) it
can be shown that the analytic expression of this overlap
is \cite{gar97}
\begin{eqnarray}
M^{JM}_{s_{12} \sigma_{12}, s_3 \sigma_3}(\bd{p}^\prime_x,\bd{p}_3) &=&
{2 \over \pi} \sum_{\ell_{12} m_{\ell_{12}} \ell_3 m_{\ell_3}}
\sum_{j_{12} L S} 
I_{\ell_{12} s_{12} j_{12}}^{\ell_3 L S}(\kappa, \alpha_\kappa)
\nonumber \\ && \hspace*{-3.5cm}
Y_{\ell_{12}m_{\ell_{12}}}(\Omega_{p^\prime_x})
Y_{\ell_3 m_{\ell_3}}(\Omega_{p_3})
\sum_{m_{12} j_3 m_3} (-1)^{J+2S-2M+\ell_3+s_3-s_{12}-\ell_{12}}
\nonumber \\ && \hspace*{-3.5cm}
\hat{j_{12}}^2 \hat{j_3}^2 \hat{J} \hat{L} \hat{S}
\left(
\begin{array}{ccc}
J& j_{12}& j_3 \\ M& -m_{12}& -m_3
\end{array}
\right)
\left(
\begin{array}{ccc}
j_3& \ell_3& s_3 \\ -m_3& m_{\ell_3}& \sigma_3
\end{array}
\right)
\nonumber \\ && \hspace*{-3.5cm}
\left(
\begin{array}{ccc}
j_{12}& \ell_{12}& s_{12} \\ -m_{12}& m_{\ell_{12}}& \sigma_{12}
\end{array}
\right)
\left\{
\begin{array}{ccc}
J& j_{12}& j_3 \\ L& \ell_{12} & \ell_3 \\ S & s_{12}& s_3
\end{array}
\right\}
\label{overl}
\end{eqnarray}
where $\kappa^2=k_x^{\prime 2}+k_y^{\prime 2}$ and 
$\alpha_\kappa=\arctan(k^\prime_x/k^\prime_y)$ 
$k^\prime_x=\sqrt{m/\mu_{12}} p^\prime_x$, 
$k^\prime_y=\sqrt{m/\mu_{12,3}} p_3$, 
$\mu_{12,3}=(m_1+m_2) m_3/(m_1+m_2+m_3)$, and $m$ is a normalization mass. 
$\ell_3$ is the relative orbital angular momentum between
the participant particle 3 and the center of mass of the spectator 
system 1+2, and it couples to $s_3$ to give
the angular momenta $j_3$. The $I$--functions are given 
in eq.(16) of \cite{gar97}, and they are computed numerically.

Inserting this into (\ref{spec}) we obtain for the spectral function
the following analytic expression:
\begin{eqnarray}
S(E^\prime_x, \bd{p}_3)& = &
\frac{p^\prime_x \mu_{12}}{\pi^3}
\sum_{j_{12} \ell_{12} s_{12} \ell_3 j_3} \sum_{L S L^\prime S^\prime}
\hat{j_{12}}^2 \hat{j_3}^2 \hat{L} \hat{S} \hat{L^\prime} \hat{S^\prime}
\label{spec2} \\ && \hspace*{-2cm}
I_{\ell_{12} s_{12} j_{12}}^{\ell_3 L S}(\kappa, \alpha_\kappa)
I_{\ell_{12} s_{12} j_{12}}^{\ell_3 L^\prime S^\prime}(\kappa, \alpha_\kappa)
\left\{
\begin{array}{ccc}
\!\!\! J& j_{12}& j_3 \!\!\! \\ 
\!\!\! L& \ell_{12} & \ell_3 \!\!\! \\ 
\!\!\! S & s_{12}& s_3 \!\!\! 
\end{array}
\right\}
\left\{
\begin{array}{ccc}
\!\!\! J& j_{12}& j_3 \!\!\! \\ 
\!\!\! L^\prime & \ell_{12} & \ell_3 \!\!\! \\ 
\!\!\! S^\prime & s_{12}& s_3 \!\!\! 
\end{array}
\right\} \nonumber
\end{eqnarray}
 
Note that the spectral function does not depend on the direction of
$\bd{p}_3$.

\subsection{Interactions}

To obtain the intrinsic wave functions (\ref{tot}) and (\ref{2beq})
the interactions between the constituents of the halo nucleus have
to be specified.

As indicated in \cite{zhu93} the details of the radial shape of
the neutron--neutron interaction are not very relevant for the
$^6$He ground state wave function as long as the low energy $n-n$
scattering parameters are correct. We then use a simple potential
that reproduces the experimental $s$--wave and $p$-wave scattering 
lengths and effective ranges. In particular we choose a potential
with gaussian shape including a central, spin--orbit
($\bd{\ell}_{nn} \cdot \bd{s}_{nn}$), tensor ($ S_{12}$), and
spin--spin ($\bd{s_1} \cdot \bd{s}_2$) interactions. This potential
was derived in \cite{cob97b}, and it is given by
\begin{eqnarray}
     V_{nn}(r)=37.05 \exp(-(r/1.31)^2)
	-7.38\exp(-(r/1.84)^2) \nonumber \\
	-23.77\exp(-(r/1.45)^2){\bd{\ell}_{nn} \cdot \bd{s}_{nn}}
	+7.16\exp(-(r/2.43)^2) S_{12}\nonumber \\+
 \left(49.40\exp(-(r/1.31)^2)+
  29.53\exp(-(r/1.84)^2) \right) {\bf s}_1 \cdot {\bf s}_2,
\end{eqnarray}
where $\bd{\ell}_{nn}$ is the relative neutron--neutron orbital angular
momentum and $\bd{s}_{nn}=\bd{s}_1 + \bd{s}_2$.

For the neutron--$\alpha$ potential we take central and spin--orbit
parts with gaussian shapes
\begin{equation}
V_{n\alpha}^{(\ell)}(r)=V_c^{(\ell)}(r) + V_{so}^{(\ell)}(r)
\bd{\ell}_{n\alpha} \cdot \bd{s}_n
\end{equation}  
where $\bd{s}_n$ is the spin of the neutron and $\bd{\ell}_{n\alpha}$
is the relative neutron--$\alpha$ orbital angular momentum.
The parameters of the gaussians are fitted to reproduce the phase
shifts for $s$, $p$, and $d$ waves up to 20 MeV. The gaussians
used are \cite{cob97} $48.2 \exp{(-r^2/2.33^2)}$ for the central $s$-wave
potential, $-47.4 \exp{(-r^2/2.30^2)}$ for the central $p$--wave potential, 
and $-21.93 \exp{(-r^2/2.03^2)}$ for the central  $d$--wave potential. 
For the spin--orbit interaction we take $-25.49 \exp{(-r^2/1.72^2)}$ for all
the waves.  All the strengths are given in MeV and the ranges in fm.
Together with the phase shifts this
interaction reproduces the $s$--wave scattering length of $-2.13$ fm,
and the two $p$--resonance energies and widths $E(p_{3/2})=0.77$ MeV,
$\Gamma(p_{3/2})=0.64$ MeV,  $E(p_{1/2})=1.97$ MeV, and 
$\Gamma(p_{1/2})=5.22$ MeV (see \cite{ajz88}).

Note that for the central $s$--wave interaction we use a repulsive
potential. This is done to avoid the neutrons from the halo to occupy
the $s$--states already occupied by the neutrons from the $\alpha$--particle,
and therefore forbidden by the Pauli principle. One could in principle
have used an attractive potential and directly exclude in the
calculation the Pauli forbidden states. As far as the repulsive
and attractive potentials have the same low--energy properties 
they give almost indistinguishable wave functions \cite{gar97b}.

It is well known as a general fact that two--body interactions 
accurately reproducing neutron--neutron and neutron--$\alpha$ low 
energy scattering data, systematically underbind the
$^6$He system by roughly 500 keV \cite{cob97}. To alleviate this problem the
range of all the interactions is normally increased by a few percent and
the central $p$--wave potential is made a bit deeper \cite{zhu93,dan93}.
However this procedure modifies the energy of the $p$--resonances,
and many observables, as momentum distributions after fragmentation
or invariant mass spectra, are very sensitive to the position of these
resonances \cite{gar97,gar97b}. We have then preferred to maintain
untouched the two--body interactions and introduce an effective
three--body force in eq.(\ref{rad}) for fine tuning.
The idea is that the three--body force should account for the polarization
of the particles beyond that described by the two--body interactions.
All the three particles must be close to produce this additional polarization,
and therefore the three--body force must be of short range. In particular
we use the attractive gaussian $-7.55 \exp{(-\rho^2/2.9^2)}$ \cite{cob97}.

The two-body neutron--neutron and neutron--$\alpha$ interactions, together
with the effective three-body force, provide after solving the Faddeev
equations a $^6$He binding energy of 0.95 MeV and a r.m.s. radius of 2.5 fm, 
in good agreement with the experimental data of $0.97\pm0.04$ MeV and 
$2.57\pm 0.10$ fm, respectively. These results are obtained including $s$,
$p$, and $d$ waves in the calculation. Three terms in the expansion
(\ref{tot}) were included. However 99.18\% of the norm of $\Psi^{JM}$
was found to be given by the first term with $n=1$, 0.75\% is given
by the second term with $n=2$, and only 0.07\% is given by the term
with $n=3$. Therefore only the first two terms are really needed in 
order to obtain an accurate enough $^6$He wave function. Exclusion of
the $n=3$--term does not modify the binding energy and r.m.s. radius. 

When the participant particle 3 is the $\alpha$--particle  the 
$\bd{x}$ coordinate connects the two spectator neutrons, and $\ell_{12}$ 
is its relative orbital momentum. In the same way $\bd{y}$ connects the
center of mass of the two neutrons and the $\alpha$, and $\ell_3$ is
its relative momentum. After computing the three--body halo wave function 
$\Psi^{JM}$ in terms of these ($\bd{x}, \bd{y}$) Jacobi coordinates it
is seen that 87.5\% of the norm is given by the $s$-waves, 9.9\% by
the $p$--waves, and 2.6\% by the $d$--waves. Therefore the relative
momentum between the two neutrons in $^6$He is basically an $s$--state,
and the $d$--waves play a minor role.
In the same way we can consider that one of the neutrons is the 
participant particle. In this case it is convenient to write $\Psi^{JM}$
in terms of the ($\bd{x}, \bd{y}$) coordinates with  $\bd{x}$ connecting
the spectator neutron and the $\alpha$--particle, and $\bd{y}$ connecting
the participant neutron and the center of mass of the spectator system
$n+\alpha$. Doing this one sees that 10.8\% of the norm  is given
by the $s$--waves, 88.5\% by the $p$--waves, and 0.7\% by the $d$--waves.
Therefore the neutron and the $\alpha$ are mainly in a relative
$p$--state. To be more precise 82.3\% of the norm is given by the
$p_{3/2}$--wave, and 6.2\% by the $p_{1/2}$. All these data are
summarized in table \ref{tab1}.

Due to the low $d$--wave content in the $^6$He wave function only $s$
and $p$ waves will be considered in the computations. The inclusion
of the $d$--waves does not produce visible changes in the results.

\begin{table}[b]
\caption{Contribution to the norm of the $^6$He wave function from the $s$, $p$,
and $d$ waves with the interactions given in the text. The second column
gives the contributions when the $\bd{x}$ coordinate connects the two neutrons
and $\bd{y}$ goes from the center of mass of the two neutrons to the
$\alpha$--particle. The third column gives the contributions when the $\bd{x}$
coordinate connects one neutron and the $\alpha$--particle
and $\bd{y}$ goes from the center of mass of the $n+\alpha$ system to the
second neutron.}
\begin{tabular}{ccc}
& \put(-42,3){\circle*{7}} \put(-12,3){\circle*{7}}
\put(-42,3){\vector(1,0){27}}
\put(-52,3){n} \put(-8,3){n} \put(-32,5){$\bd{x}$}
& \put(-42,3){\circle*{13}} \put(-12,3){\circle*{7}}
\put(-42,3){\vector(1,0){27}}
\put(-56,3){$\alpha$} \put(-8,3){n} \put(-29,5){$\bd{x}$} \\
$\ell_{12}=0$ & \hspace*{-2cm} 87.5\% & \hspace*{-2cm} 10.8\% \\
$\ell_{12}=1$ & \hspace*{-2cm} 9.9\%  & \hspace*{-2cm} 88.5\%\\
$\ell_{12}=2$ & \hspace*{-2cm} 2.6\%  & \hspace*{-2cm} 0.7\%
\end{tabular}
\label{tab1}
\end{table}

\end{multicols}
\begin{figure}[ht]
\centerline{\psfig{figure=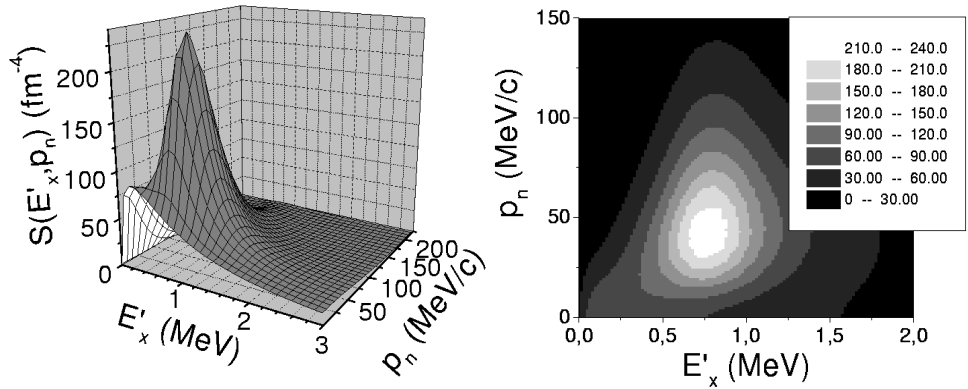,width=8.5cm,%
bbllx=5.7cm,bblly=6.5cm,bburx=15.2cm,bbury=21.2cm,angle=270}}
\caption[]{
Spectral function (eq.(\ref{spec})) after knockout of one
of the halo neutrons in $^6$He by the electron. The figure in the right
side is
the contour plot of the spectral function shown in the left part.
}
\end{figure}
\begin{multicols}{2}

\section{Results}
 
In this section we give the results for the observables derived in 
section IV for an electron scattering process on $^6$He.  The
computations are done following the steps given in section V.

All the observables can be computed in two different scenarios. First,
assuming that the participant particle is one of the neutrons and the
unbound $^5$He system survives in the final state; and second, the 
$\alpha$ particle is the one knocked out by the electron and the two
neutrons survive undisturbed in the final state. In both cases we
consider that we are in the corresponding quasielastic peak, 
i.e. the energy transfer is $\omega=q^2/2M$ where $M$ is the neutron 
mass in the first case and the $^4$He mass in the second case.

\subsection{Spectral functions and momentum distributions}

   The spectral function $S(E^\prime_x, p_3)$ is given in (\ref{spec}),
while in (\ref{spec2}) we show its analytic form obtained from
the specific expressions used for the wave functions. As already 
mentioned, this function gives the probability of removing the 
participant particle with momentum $p_3$ leaving the residual 
system with energy $E^\prime_x$.

\end{multicols}
\begin{figure}[ht]
\centerline{\psfig{figure=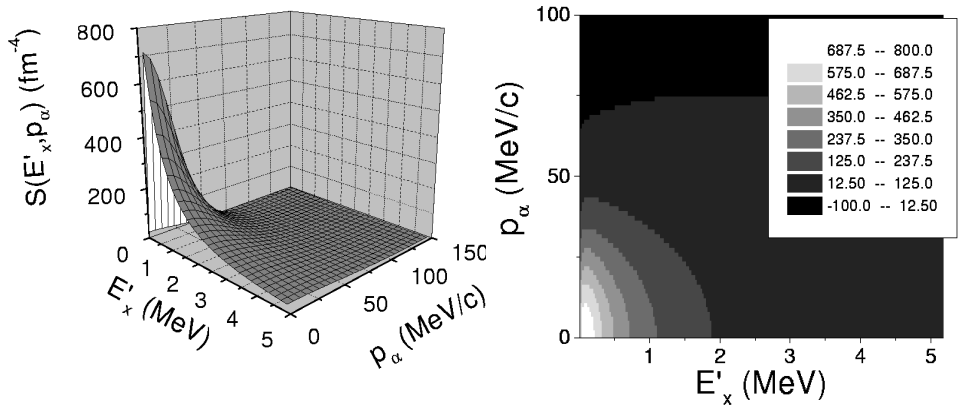,width=8.5cm,%
bbllx=5.5cm,bblly=6.5cm,bburx=15.2cm,bbury=21.2cm,angle=270}}
\caption[]{
Same as fig.4 for the case of $\alpha$ knockout by
the electron.
}
\end{figure}
\begin{multicols}{2}

  Let us start with the case of a participant neutron. The spectral function
is shown in fig.4, and reflects the momentum distribution of the internal 
nucleon and the energy spectrum of the unbound system $^5$He. In the 
left part of the figure we show the spectral function itself, while the 
right part shows the projection over the basis.  

  The spectral function has a peak whose precise position is easily seen in 
the right part of the figure. The maximum value  corresponds to a $^5$He
energy of roughly $E^\prime_x=0.75$ MeV, that is the energy of the lowest
$p_{3/2}$ resonance in $^5$He (0.77 MeV \cite{ajz88}). The width of this 
resonance is relatively small (less than 0.7 MeV), and this makes the peak 
clearly pronounced. The next resonance in $^5$He ($p_{1/2}$) has an energy 
of around 2 MeV, and it should also appear in the figure. However, its large 
width, more than 5 MeV, makes this resonance to be smeared out and hidden
by the $p_{3/2}$. Furthermore, as mentioned in subsection V.D,
more than 82\% of the norm of the $^6$He wave
function is given by the $p_{3/2}$ wave, while the $p_{1/2}$ only gives around
6\%. The presence of $s$--waves in the $^6$He wave function
($\sim$ 10\%) makes the spectral function non zero along the $p_n=0$ edge.

  In fig.5 we show the same as in fig.4 but for the case where the 
$\alpha$--particle is scattered away by the electron, and the two neutrons
act as spectators. In this case the overlap (\ref{overl}) is computed
writing the $^6$He wave function in terms of the Jacobi coordinates
with $\bd{x}$ connecting both neutrons. From table \ref{tab1} we know
that both neutrons are preferably in a relative $s$--state, and this
will produce a non zero spectral function along the $p_\alpha=0$
axis. The fact that the spectral function is proportional to 
$\sqrt{E^\prime_x}$ (see eq.(\ref{spec2})) will make the function
equal to zero at the origin. 
This peaked spectral function close
to the origin reflects the fact that the $n-n$ interaction has
a very large scattering length for the $^1S_0$--wave (more than 23 fm),
giving rise to a virtual state at very low energy (less than 100 keV).

\begin{figure}[ht]
\centerline{\psfig{figure=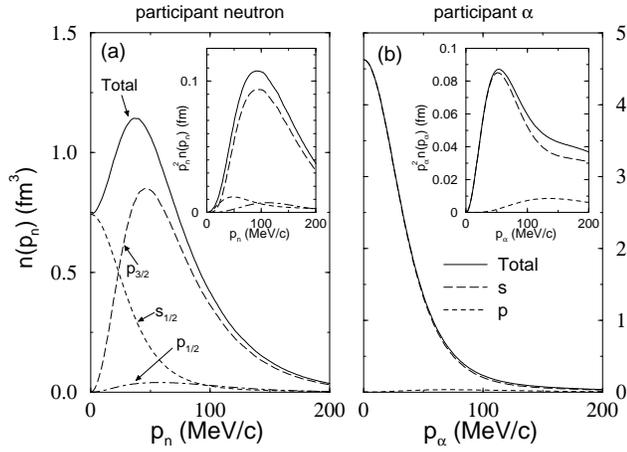,width=5cm,%
bbllx=3.cm,bblly=6.5cm,bburx=15.2cm,bbury=21.2cm,angle=270}}
\vspace{1.9cm}
\caption[]{
(a) Neutron momentum distribution (see eq.(\ref{eq35}))
for the participant neutron case. The contributions to the total from the
$s_{1/2}$
wave (short--dashed), $p_{1/2}$ wave (dot--dashed), and $p_{3/2}$ wave
(long--dashed) are shown. The inset shows the same contributions multiplied 
by $p_n^2$,
where the weight of each wave can be better appreciated. (b) Same as
(a) for the $\alpha$ momentum distribution in the participant $\alpha$
case.
}
\end{figure}

After integration of the spectral function over the internal two--body
energy $E^\prime_x$ we obtain the internal momentum distribution 
of the  participant particle (eq.(\ref{eq35})). In fig.6a
we show the computed neutron  momentum distribution after integration
over $E^\prime_x$ of the spectral function in fig.4 (solid line). The
contributions to the total from the $s_{1/2}$ wave (short--dashed line),
$p_{1/2}$ wave (dot--dashed line), and $p_{3/2}$ wave (long--dashed line)
are also shown. It may look that the $s$--wave contribution is much larger than
the 10\% given in table I. This is because the whole contribution at 
the origin comes from the $s$ waves. However, when computing the normalization
of the momentum distribution the phase volume contains a $p_n^2$ factor,
that reduces the weight of the $s$ contribution. In the inset of fig.6a
we show the neutron momentum distribution multiplied by $p_n^2$. We
clearly see that dominant contribution comes from the $p_{3/2}$ wave,
while the $s$ contribution is small.

In fig.6b we show the same but for the case of participant $\alpha$.
Again the solid line shows the total $\alpha$ momentum distribution, and
the long--dashed and short--dashed correspond to the contributions from
the $s$ and $p$ waves, respectively. In this case (table \ref{tab1}) the 
dominant term is the relative neutron--neutron state with orbital angular 
momentum zero. In fact the long--dashed curve and the solid one coincide
almost in the whole range of $\alpha$ momenta. Even when the momentum
distribution is multiplied by the $p_\alpha^2$ factor coming from
the phase volume (inset in fig.6b) it is seen that the $s$ waves
are the clearly dominant ones, and they are responsible for the
pronounced peak observed in the spectral function in fig.5.

\subsection{Cross sections and response functions}

To compute the cross sections (\ref{dif5}) and (\ref{dif5b}) we 
work in the {\it in plane} perpendicular kinematics (see end of
section IV). The momentum transfer $q$ is fixed at a value of
$q=190$ MeV/c, and the energy transfer corresponds
to the center of the quasielastic peaks ($\omega=q^2/2m_3$), with
$m_3=m_n$ or $m_3=m_\alpha$.
The cross sections (\ref{dif5}) and (\ref{dif5b}) are then functions
of the polar angles of the particle detected in coincidence with
the electron, i.e. $\theta_{p^\prime_3}$ and  $\theta_{p^\prime_i}$,
respectively.

{\it i) Participant neutron.}

Let us first consider the case when $\omega$ corresponds to the center 
of the neutron quasielastic peak. As discussed at the end of section II 
only a halo
neutron can be the participant particle. If the particle detected in
coincidence with the electron is the participant neutron ($(e,e^\prime n)$
process) the cross section is given by eq.(\ref{dif5}). Assuming that
the internal energy $E^\prime_x$ can be neglected in eq.(\ref{delta}) the
cross section can then also be written as shown in (\ref{fincross}).

In fig.7 we show the differential cross section for this process. Two
cases have been considered, one of them at a forward scattering angle
($\theta_e=30$), and the second one at a backward scattering angle 
($\theta_e=150$). The electron beam energy $E_0$ takes
a value of 375 MeV in the first case, and 108 MeV in the second.
In the figure the solid line is the calculation as given in eq.(\ref{dif5}).
The dashed line is the calculation as in (\ref{fincross}) where $E^\prime_x$
has been fully neglected in eq.(\ref{delta}), i.e. $E^\prime_x=0$.
Finally the long dashed line is the calculation as in (\ref{fincross}) but
taking $E^\prime_x=0.77$ MeV in eq.(\ref{delta}). This energy corresponds
to the energy of the lowest $p$--resonance in $^5$He.

\begin{figure}[ht]
\centerline{\psfig{figure=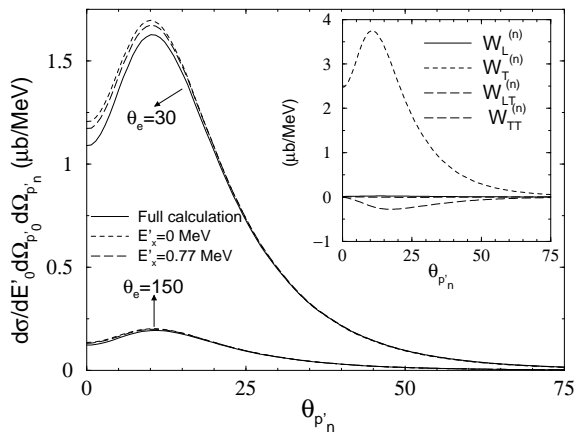,width=5cm,%
bbllx=3.8cm,bblly=6.5cm,bburx=15.2cm,bbury=21.2cm,angle=270}}
\vspace{1.9cm}
\caption[]{
Differential cross section for a $(e,e^\prime n)$ reaction
for a participant neutron process. The participant neutron is assumed to be
detected in coincidence with the electron. The calculations with scattering
angles
$\theta_e=30$ and $\theta_e=150$ are shown. The solid lines are calculations
as given in (\ref{dif5}). The short--dashed and long--dashed lines
are calculations as given in (\ref{dif5b}) taking $E^\prime_x=0$
and $E^\prime_x=0.77$ MeV in (\ref{delta}), respectively. The inset
shows the structure functions appearing in (\ref{dif5}).
}
\end{figure}

As seen in the figure the three computed cross sections are very similar,
with only a small difference in the maximum, that is placed at a value
of $\theta_{p^\prime_n}$ of roughly 10 degrees. When the approximation
in (\ref{fincross}) is used the chosen value for $E^\prime_x$ in eq.(\ref{delta})
is not very relevant, and only a full calculation as in (\ref{dif5}) reduces
the maximum and the value at $\theta_{p^\prime_n}=0$. In all cases 
the cross section is already negligible for outgoing neutrons with polar 
angle larger than 75 degrees. 

The calculations with different scattering angles differ roughly in a global 
scale factor. This is because for a participant neutron only the magnetic
nucleon form factor is important. Hence, in eq.(\ref{fincross}) the
main contribution is given by the transverse structure function, and the
whole dependence on the scattering angle is contained in the global
factor $\sigma_M V_T$. From eqs.(\ref{smott}) and (\ref{vr}) one sees that
$\sigma_M V_T$ goes like $1/\sin^2(\theta_e/2)$ and scattering angles
close to zero give a much larger contribution than scattering angles
close to $\pi$. This is seen in fig.7, where the maximum value of
the cross section is 8 times larger for $\theta_e=30$ than for 
$\theta_e=150$. In the inset in fig.7 we show the four response functions
in eq.(\ref{fincross}). It is clear that, as mentioned above, the 
dominant contribution is by far given by the 
transverse part. Only $W_{LT}$ gives an additional non negligible 
contribution. 

When the participant particle  is one of the halo neutrons,
either the second halo neutron or the core can also be detected in 
coincidence with the electron. In this case the cross section is given
by eq.(\ref{dif5b}), where $p^\prime_i$ is the momentum either of the
spectator neutron or of the spectator core (eq.(\ref{surv})). The solid lines
in fig.8a give the differential cross section when the particle detected
in coincidence with the electron is the second neutron.
Therefore it also contributes to the $e^\prime n$ coincidence cross 
section. The two cases with
$\theta_e=30$ (thick lines) and $\theta_e=150$ (thin lines) are shown. 
As seen in the figure, for both scattering angles 
the maximum contribution to the cross section appears at an angle
of around 70 degrees, value for which the cross sections shown in fig.7
are already small. Furthermore for small angles the contribution to
the $e^\prime n$ coincidence cross section coming from the spectator neutron
is more than 10 times smaller than the contribution from the participant
neutron (see fig.7). In fig.8b we show the total differential cross
section (solid line) of the $e^\prime n$ coincidence reaction for $\theta_e=30$. 
This curve is obtained by summing up the contribution from the participant
neutron (short--dashed line in fig.8b, or solid line for $\theta_e=30$ in fig.7)
and the contribution from the spectator neutron (thick solid line in fig.8a).
From fig.8b it is then
clear that the behavior of the cross section is dominated by
the participant neutron, while the spectator neutron contributes
with a roughly constant cross section that creates a long tail that 
extends all the way up to $\theta_{p^\prime_n}=180$.
When $\theta_e=150$ 
the difference is basically a global factor close to 8.
We also show in fig.8a the differential cross section for the
$e^\prime \alpha$ coincidence process when the $\alpha$ particle is
a spectator. They are given by the dashed curves, and show a pronounced 
peak at 70 degrees. 

\begin{figure}[ht]
\centerline{\psfig{figure=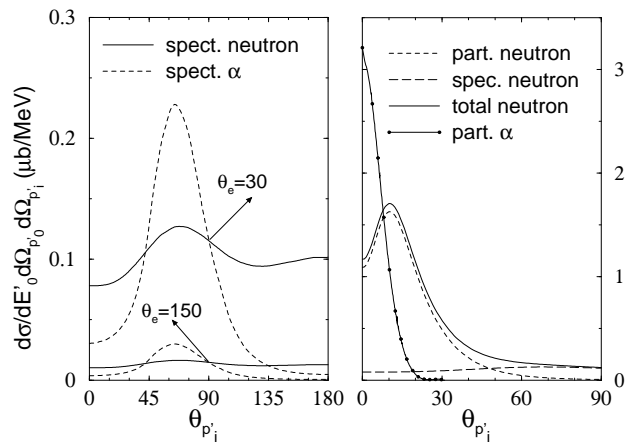,width=5cm,%
bbllx=3.7cm,bblly=6.5cm,bburx=15.2cm,bbury=21.2cm,angle=270}}
\vspace{1.9cm}
\caption[]{
(a) Differential cross sections for coincidence reactions
after a participant neutron process and assuming that the particle
detected in coincidence with the electron is either the spectator neutron
(solid line) or the spectator $\alpha$ (dashed line). Calculations
with $\theta_e=30$ (thick lines) and $\theta_e=150$ (think lines) are
shown. (b) Total differential cross section after an electron--neutron
coincidence reaction (solid line). The contributions from the
participant neutron (short--dashed line) and the spectator neutron
(long--dashed line) are shown. The dotted line is the differential cross
section for a $(e,e^\prime \alpha)$ process assuming that the $\alpha$ is the
participant particle.
}
\end{figure}

{\it ii) Participant $\alpha$.}

Let us finish with the case when $\omega$ corresponds to the center
of the $\alpha$ quasielastic peak. The participant particle is
then the $\alpha$, $m_3=m_\alpha$, and for $q=190$ MeV/c the energy
transfer is around 5 MeV. Neglecting $E^\prime_x$ in eq.(\ref{delta}),
and taking $M_h=m_1+m_2+m_3$, it can be seen that the angle $\theta_{p^\prime_3}$
has to satisfy that $\sin \theta_{p^\prime_3} \leq (m_1+m_2)/m_3$. This
relation is fulfilled for any angle when the participant particle is
one of the halo neutrons, but when $m_3 > (m_1+m_2)$ it gives a limit
to the value of $\theta_{p^\prime_3}$. In particular, for a 
participant $\alpha$ particle one has that $\theta_{p^\prime_3} \lesssim 30$. 
This is also shown in fig.8b, where the dotted curve shows the differential cross 
section for an exclusive $(e,e^\prime \alpha)$ reaction in the center 
of the $\alpha$ quasielastic peak. The cross section is zero for angles
larger than 30, and the maximum value at the origin is around
twice the one obtained in the $e^\prime n$ coincidence process. 
In the computation
shown in the figure the scattering angle is 30 degrees, 
for which $E_0=369$ MeV ($q=190$ MeV/c, $\omega=q^2/2m_\alpha$).
As seen in eq.(\ref{alpha}), 
only a longitudinal structure function enters in this case, and
the whole dependence on the scattering angle is then contained
in $\sigma_M$. Therefore different scattering angles only introduce 
a global factor determined by
the ratio between the different values of the Mott cross section.
For instance, a scattering angle of 15 degrees would make the 
differential cross section four times larger than the one shown
in the figure, while $\theta_e=150$ would make the cross section
more than 200 times smaller than the one in the figure. 
After $\alpha$ knockout by the electron the spectator neutrons
could in principle be detected in coincidence with the outgoing 
electron. The computed cross section for this process is very
small, and barely visible in the scale of fig.8b.

\section{Summary and final remarks}

The fact that electron scattering is one of the most powerful and
cleanest procedures in order to investigate nuclear structure lead
us to consider these reactions to investigate one of the most intriguing 
objects recently discovered in nuclear physics: Halo nuclei. In 
particular we have considered the case of two--neutron halo nuclei.
The three--body  wave function of the nucleus is obtained by solving
the Faddeev equations in coordinate space by means of the adiabatic 
hyperspherical expansion. As soon as the two--body interactions
between the constituents of the nucleus are specified the three--body
structure can be computed to the needed accuracy.  

First we have made a short incursion into elastic electron scattering reactions 
where we have seen the variation of the charge form factor according to
two limiting values for the charge r.m.s. radius, the core size and the
$^6$He size.  The validity of the three--body picture requires that
the charge form factor for $^6$He and $^4$He should be similar. In other
words, if the present picture of $^6$He as a $^4$He core and two halo 
neutrons is correct, the elastic differential cross section for $^6$He
should be practically identical to that of $^4$He, while if the spatial
charge distribution of $^6$He would be as extended as that of the
neutrons, the cross section would be at least four times smaller at
typical $q$ values of 200 -- 300 MeV/c. Thus elastic electron scattering
provides a direct test of the halo picture.

 Then we make an extensive discussion of $(e, e^\prime x)$
processes, that are described in the impulse approximation, i.e.
the virtual photon is absorbed by a single constituent of the
halo nucleus. The quasielastic exclusive electron scattering formalism 
was developed years ago (see for instance \cite{fru84}), and in principle 
one could directly apply it to electron scattering on halo nuclei. 
However, the language used in the three--body description of the halo 
nucleus (inert core, Jacobi coordinates, ...) is not easily matched with 
the language commonly used in electron scattering, where the single--particle 
properties of nuclei play an important role. 
One of the main aims of this paper has been  to obtain the differential cross
section for electron scattering using the three--body  wave function
of the halo nucleus as starting point. Comparison of the derived cross 
section  with the well known expression for quasielastic exclusive electron 
scattering permits an easy  identification of the observables of interest: 
Spectral function, response functions, and momentum distributions. 

After discussion of the kinematics of the reaction we have applied
the method to electron scattering on $^6$He, for which the two--body 
interactions between its three constituents are well known. These 
interactions reproduce the basic properties of the nucleus, separation energy 
and root mean square radius, together with  an important amount of experimental 
data obtained after fragmentation on stable targets. Uncertainties coming
from a poor description of the halo nucleus are then minimized. We have 
investigated the different observables for the cases of a participant neutron  
and a participant $\alpha$.

We have first studied the spectral function, that  carries the whole 
information about the structure of the two--body subsystem obtained after 
removal by the electron of one of the constituents of $^6$He. This 
function is interpreted as the probability of removing one of the
constituents with a certain momentum leaving the residual system with
a certain energy. Effects due to the presence of different partial waves 
can be observed, as well as indications about the resonance structure of the 
surviving unbound two--body subsystem.  Of special interest
is the case of neutron removal, since the spectral function directly
gives information about the unbound nucleus $^5$He.

After integration of the spectral function over the energy we obtain the 
momentum distribution of the participant particle, that contains 
information about the weight of each partial wave in the wave function.
Experimental information about the momentum distribution can in 
principle be obtained by taking the ratio between 
the measured cross section and the 
electron--participant particle cross section. However, one has to keep
in mind that the factorization of the five--differential cross section
is only approximated, and low lying resonances in the final unbound 
two--body system are required. 
For the same reason as before, low lying resonances in the two--body 
system are necessary in order to write the five--differential cross section
in terms of the response functions (that are simply the product of the momentum 
distribution and the corresponding single--particle response function).
This can always be done when the residual nuclear system is bound. We have seen 
that for $^5$He (that has a resonance at an energy of around 0.8 MeV) the 
approximated differential cross section does not differ very much from the exact
one, and it is therefore possible to talk about response functions in this 
kind of reactions. Such a response functions have 
been obtained for the case of a participant neutron, where the transverse
response clearly dominates, and for a participant $\alpha$, where only the
longitudinal one enters. 

In summary, as for any nucleus, electron scattering is probably the most
accurate procedure in order to investigate the structure of halo nuclei.
Due to the up to now unattainable technical problems, experimental
measurements are not available, and these reactions have been completely
unexplored. Fortunately the first experiments of this kind are projected
for the next few years \cite{ohk97}, and theoretical studies 
as the ones presented here are then needed. We have derived 
the cross section  and investigated for the case of the two--neutron
halo nucleus $^6$He how the observables are connected to the structure of the 
nucleus. We show results for differential cross sections corresponding
to elastic scattering as well as to quasielastic scattering on a neutron halo 
and on the $^4$He core,
illustrating how the different kinematical regions probe the charge distribution,
the spectral function and the momentum distributions in a transparent 
manner. Additional information could also be obtained exploiting
spin polarization degrees of freedom, which are not discussed here 
given the incipient state of the subject.

\acknowledgments

We are thankful to A.S. Jensen for useful comments and suggestions.
This work was supported by DGICYT (Spain) under contract number PB95/0123.

\appendix 
\section{Derivation of the transition matrix}
In fig.2 we show the scheme of the exclusive $(e,e^\prime x)$ reaction 
together with the notation for the different energies and momenta
involved. The axis system chosen to describe the process is given 
in fig.1b. 
If we denote by $\bd{r}_i$ ($i=0,1,2,3$) the position
of the electron and the three particles in the halo nucleus we
then define the coordinates:
\begin{eqnarray}
\bd{x}&=&\bd{r}_1-\bd{r}_2 \nonumber \\
\bd{y}&=&\frac{m_1 \bd{r}_1 + m_2 \bd{r}_2}{m_1+m_2} - \bd{r}_3 \label{jac} \\
\bd{z}&=&\bd{r}_0-\bd{r}_3 \nonumber
\end{eqnarray}

The coordinates $\bd{x}$ and $\bd{y}$ are the Jacobi coordinates
except for the mass factors that are usually introduced and that
for simplicity have been omitted here (see for instance \cite{gar97}).

The conjugated momenta associated to the coordinates $\bd{x}$, $\bd{y}$,
and $\bd{z}$ are
\begin{eqnarray}
\bd{p}_x &=& \frac{m_2}{m_1+m_2} \bd{p}_1 - \frac{m_1}{m_1+m_2} \bd{p}_2
                                    \nonumber \\
\bd{p}_y &=& \frac{m_3}{m_1+m_2+m_3} (\bd{p}_1+\bd{p}_2) - 
\frac{m_1+m_2}{m_1+m_2+m_3} \bd{p}_3
                                    \label{mom} \\
\bd{p}_z &=& \frac{m_3}{m_0+m_3} \bd{p}_0 - \frac{m_0}{m_0+m_3} \bd{p}_3
                    \nonumber
\end{eqnarray}
and the same with primes in the final state. The masses of the three 
halo constituents are denoted by $m_i$ ($i=1,2,3$).

Working in the frame of the halo nucleus ($\bd{p}_h=
\bd{p}_1+\bd{p}_2+\bd{p}_3=0$) one has 
\begin{equation}
\bd{p}_y=\bd{p}_1+\bd{p}_2=-\bd{p}_3
\end{equation}

Following \cite{bjo64} in the one photon exchange approximation 
the transition matrix element for an electron
scattering process is given by
\begin{equation}
S_{fi}=-ie \int dt d\bd{r}_0 \bar{\psi}_f(t,\bd{r}_0) \gamma_\mu 
\psi_i(t,\bd{r}_0) A^\mu(t,\bd{r}_0)
\label{mat}
\end{equation}
where $\psi_{i,f}$ are the initial and final electron wave functions,
that are given by
\begin{equation}
\psi_i(t,\bd{r}_0)=\frac{1}{\sqrt{V}} u(\bd{p}_0,\sigma_0) e^{-iE_0t}
                 e^{i \bd{p}_0 \cdot \bd{r}_0}
\label{ini}
\end{equation}
\begin{equation}
\psi_f(t,\bd{r}_0)=\frac{1}{\sqrt{V}} u(\bd{p}^\prime_0,\sigma^\prime_0) 
          e^{-iE_0^\prime t} e^{i \bd{p}_0^\prime \cdot \bd{r}_0}
\label{fin}
\end{equation}
that are normalized to 1 in a volume $V$. $\sigma_0$ and $\sigma_0^\prime$
are the third components of the spin for the ingoing and outgoing electrons,
respectively.

The four-vector $A^\mu(t,\bd{r}_0)$ is the potential (scalar and vector)
generated by the nucleus and seen by the electron at the instant $t$
in the position $\bd{r}_0$.

Substituting (\ref{ini}) and (\ref{fin}) into (\ref{mat}) one has
\begin{eqnarray}
\lefteqn{ S_{fi}=} \label{mat2} \\ &&
\frac{-ie}{V} \bar{u}(\bd{p}^\prime_0,\sigma^\prime_0) \gamma_\mu
u(\bd{p}_0,\sigma_0) \int dt d\bd{r}_0 e^{-i \omega t} e^{i \bd{q} \cdot 
\bd{r}_0 } A^\mu(t,\bd{r}_0) 
\nonumber
\end{eqnarray}

Making use of the Maxwell equations one has 
\begin{equation}
\Box A^\mu(t,\bd{r_0}) = e \langle f | \hat{J}^\mu | i \rangle
\label{max}
\end{equation}
and the transition matrix element takes the form \cite{don75}
\begin{eqnarray}
S_{fi}&=&\frac{-ie^2}{V} \bar{u}(\bd{p}^\prime_0,\sigma^\prime_0) \gamma_\mu
u(\bd{p}_0,\sigma_0) \nonumber \\ &&
\frac{1}{q_\rho^2} \int dt d\bd{r}_0 
e^{-i \omega t} e^{i \bd{q} \cdot \bd{r}_0 }  
                      \langle f | \hat{J}^\mu | i \rangle
\label{mat3}
\end{eqnarray}
where $\hat{J}^\mu$ is the nuclear current operator 
($\hat{J}^\mu=(\hat{\rho}, \hat{\bd{J}})$). 

The initial hadronic state $|i\rangle$ is given by the three--body halo wave
function that is written as
\begin{equation}
\Psi^{JM}(\bd{x},\bd{y}) \frac{1}{\sqrt{V}} e^{-i E_h t} 
                         e^{i \bd{p}_h \cdot \bd{r}_h}
\label{3bod}
\end{equation}
with
\begin{equation}
\bd{r}_h = \frac{m_1 \bd{r}_1 + m_2 \bd{r}_2 +m_3 \bd{r}_1}{m_1+m_2+m_3}
\end{equation}
and  $\Psi^{JM}(\bd{x},\bd{y})$ is the intrinsic three--body halo 
wave function with spin $J$ and third component $M$.

The wave function of the two--body system made by particles 1 and 2
in the final state is written as
\begin{equation}
\frac{1}{V}
w^{s_{12},\sigma_{12}}(\bd{p}^\prime_x,\bd{x}) 
e^{-i (E^\prime_1+E^\prime_2) t} 
e^{i(\bd{p}^\prime_1+\bd{p}^\prime_2)\cdot \bd{r}_{12}}
\label{sys12}
\end{equation}
where $w^{s_{12},\sigma_{12}}(\bd{p}^\prime_x,\bd{x})$ is the intrinsic
continuum wave function, $s_{12}$ is the coupling of the spins of both 
particles and $\sigma_{12}$ its third component, and 
$\bd{r}_{12}=(m_1 \bd{r}_1 + m_2 \bd{r}_2 )/(m_1+m_2)$.

The final hadronic state $|f\rangle$ is made by the product of 
eq.(\ref{sys12}) and the plane wave describing the outgoing particle 3
\begin{equation}
\frac{1}{\sqrt{V}} e^{-i E^\prime_3 t} e^{i \bd{p}^\prime_3 \cdot \bd{r}_3 }
            \chi_{s_3,\sigma_3^\prime}
\label{par3}
\end{equation}
where $\chi_{s_3,\sigma_3^\prime}$ is the spin state of particle 3 in the final
state.

According to the scheme shown in fig.2 particle 3 is the only constituent
interacting with the electron. The rest of them are just spectators. 
The nuclear current operator should then be the one associated to the hadron 
labeled by 3. Assuming that this current operator depends only on the distance
$z$ between particle 3 and the electron and substituting (\ref{3bod}), 
(\ref{sys12}), and (\ref{par3}) into (\ref{mat3}) a little algebra leads
to 
\begin{eqnarray}
S_{fi}&=&\frac{i}{V^3} \frac{e^2}{q_\rho^2} (2 \pi)^4
\delta(E_i-E_f) \delta^3(\bd{P}_i-\bd{P}_f) \label{mat4} \\ &&
\bar{u}(\bd{p}^\prime_0,\sigma^\prime_0) \gamma_\mu
u(\bd{p}_0,\sigma_0) \nonumber \\ &&
\sum_{\sigma_3} \int d\bd{z} e^{i \bd{q} \cdot \bd{z}} 
\langle \chi_{s_3,\sigma_3^\prime}
|\hat{J^\mu}(\bd{z}) | \chi_{s_3,\sigma_3} \rangle
(2 \pi)^3 \nonumber \\ &&
\langle \frac{1}{(2 \pi)^3} e^{i \bd{p}_y \cdot \bd{y} }
\chi_{s_3,\sigma_3}
w^{s_{12},\sigma_{12}}(\bd{x},\bd{p^\prime_x}) | 
\Psi^{JM}(\bd{x},\bd{y}) \rangle
\nonumber
\end{eqnarray}
where we have used that 
$d\bd{r}_0d\bd{r}_1d\bd{r}_2d\bd{r}_3=d\bd{r}_0d\bd{x}d\bd{y}d\bd{z}$,
and where the integrations over $t$ and $\bd{r}_0$ give rise to
the deltas forcing energy and momentum conservation. $E_i=E_0+E_h$
and $E_f=E^\prime_0+E^\prime_1+E^\prime_2+E^\prime_3$ are the initial
and final total energies, and analogously for the initial and
final total momenta $\bd{P}_i$ and $\bd{P}_f$.

In deriving (\ref{mat4}) we have also used that
particle 3 absorbs the whole energy and momentum transfer to
the nucleus, and then in the frame of the three--body halo system one
has
\begin{equation}
\bd{p}_1+\bd{p}_2=\bd{p}_y=-\bd{p}_3=\bd{p}^\prime_1 + \bd{p}^\prime_2
=\bd{p}^\prime_y
\label{relat}
\end{equation}
Therefore in this picture the center of mass momentum of the final 
two--body system made by particles 1 and 2 reveals the internal momentum
of the participant particle 3.

We denote now 
\begin{equation}
J^\mu_{\sigma_3^\prime,\sigma_3}(\bd{q},\bd{p}_3) = \frac{1}{(2 \pi)^{3/2}}
\int d\bd{z} e^{i \bd{q} \cdot \bd{z}} \langle \chi_{s_3,\sigma_3^\prime}
|\hat{J^\mu}(\bd{z}) | \chi_{s_3,\sigma_3} \rangle
\label{cur}
\end{equation}
that is the matrix element in momentum space of the current operator
connecting states of particle 3 with spin projections $\sigma_3$
and $\sigma^\prime_3$.

We also define
\begin{eqnarray}
\lefteqn{
M^{JM}_{s_{12} \sigma_{12}, s_3 \sigma_3}(\bd{p}^\prime_x,\bd{p}^\prime_y)
=} \nonumber \\ &&
\langle \frac{1}{(2 \pi)^3} e^{i \bd{p}^\prime_y \cdot \bd{y} }
\chi_{s_3,\sigma_3}
w^{s_{12},\sigma_{12}}(\bd{x},\bd{p^\prime_x}) |
\Psi^{JM}(\bd{x},\bd{y}) \rangle
\label{sudd}
\end{eqnarray}
that is the overlap between the initial three--body halo wave 
function and the wave function of the final two--body system
(the exponential is the two--body center of mass motion). When
the interaction between particles 1 and 2 is neglected the distorted
function $w^{s_{12},\sigma_{12}}(\bd{x},\bd{p^\prime_x})$ becomes
a plane wave ($e^{i \bd{p^\prime_x} \cdot \bd{x}} \chi_{s_{12},\sigma_{12}}$)
and (\ref{sudd}) is the Fourier transform (normalized to 1) of the
three-body halo wave function.

With (\ref{cur}) and (\ref{sudd}) we can finally write the transition
matrix element (\ref{mat4}) in the form
\begin{eqnarray}
S_{fi}&=&\frac{i}{V^3} \frac{e^2}{q_\rho^2} (2 \pi)^{17/2}
\delta(E_i-E_f) \delta^3(\bd{P}_i-\bd{P}_f) \nonumber \\ &&
\bar{u}(\bd{p}^\prime_0,\sigma^\prime_0) \gamma_\mu
u(\bd{p}_0,\sigma_0) \nonumber \\ && 
\sum_{\sigma_3} J^\mu_{\sigma_3^\prime,\sigma_3}(\bd{q},\bd{p}_3)
M^{JM}_{s_{12} \sigma_{12}, s_3 \sigma_3}(\bd{p}^\prime_x,\bd{p}^\prime_y)
\label{matrix}
\end{eqnarray}

\section{Electron-nucleon off--shell structure functions}
For the off--shell electron-nucleon cross section we have chosen what
in refs.\cite{for83,cab93} is called prescription CC1 with current
conservation.

Let us call
\begin{equation}
\alpha=F_1(q_\mu^2)+F_2(q_\mu^2); 
\hspace*{1.5cm}
\beta=\frac{-F_2(q_\mu^2)}{2M_{\mbox{\scriptsize nucleon}}}
\end{equation} 

The response functions defined as in eq.(\ref{vr}) are then ({\it in plane}
kinematics is assumed)
\begin{eqnarray}
{\cal R}_L&=&\frac{1}{2E_3E_3^\prime} \left[
\alpha^2 (2E_3E_3^\prime + m_3^2-p_3^\mu p_{3\mu}^\prime )+ 
\right. \nonumber \\ && \left.
(E_3+E^\prime_3)^2(\beta^2(m_3^2+p_3^\mu p_{3\mu}^\prime )+2\alpha\beta m_3)
\right]
\end{eqnarray}
\begin{eqnarray}
{\cal R}_T&=&\frac{1}{E_3E_3^\prime} \left[
\alpha^2 (p^{\prime 2}_3 \sin^2\theta_{p^\prime_3}- 
   m_3^2+p_3^\mu p_{3\mu}^\prime)+ \right. \nonumber \\ && \left.
2 p^{\prime 2}_3 \sin^2\theta_{p^\prime_3} 
(\beta^2(m_3^2+p_3^\mu p_{3\mu}^\prime )+2\alpha\beta m_3)
\right]
\end{eqnarray}
\begin{eqnarray}
{\cal R}_{LT}&=&-\frac{\sqrt{2}}{E_3E_3^\prime} p^\prime_3 \sin \theta_{p^\prime_3}
(E_3+E^\prime_3) \nonumber \\ && 
\left[
\alpha^2 + 2\beta^2 (m_3^2+p_3^\mu p_{3\mu}^\prime ) +4\alpha \beta m_3
\right]
\end{eqnarray}
\begin{eqnarray}
{\cal R}_{TT}&=&-\frac{1}{E_3E_3^\prime} p^{\prime 2}_3 \sin^2\theta_{p^\prime_3}
\nonumber \\ && \left[
\alpha^2 + 2\beta^2 (m_3^2+p_3^\mu p_{3\mu}^\prime) +4\alpha \beta m_3
\right]
\end{eqnarray}

Finally
\begin{equation}
F_1(q_\mu^2)=\frac{G_E+\tau G_M}{1+\tau};
\hspace*{1.cm}
F_2(q_\mu^2)=\frac{G_M-G_E}{1+\tau}
\end{equation}
where $\tau=-q_\mu^2/(4 M_{\mbox{\scriptsize nucleon}})$,
and $G_E$ and $G_M$ are parameterized as in \cite{gal71}.

\end{multicols}

\end{document}